\documentclass[final,1p,times,number]{elsarticle}

\usepackage{amsmath}
\usepackage{amssymb}
\usepackage{amsfonts}
\usepackage{graphicx}
\usepackage{epstopdf}
\usepackage{color}
\usepackage{bm}
\usepackage{multirow}
\usepackage{natbib}
\usepackage{hyperref}

\usepackage{ifpdf}
\ifpdf%
\usepackage{pdflscape}
\else
\usepackage{lscape}
\fi

\DeclareMathOperator{\im}{Im}

\DeclareMathOperator{\re}{Re}
\DeclareMathOperator{\arcsinh}{arcsinh}

\journal{Annals of Physics}
\biboptions{sort&compress}

\begin{document}

\begin{frontmatter}
\title{Absolute Poisson's ratio and the bending rigidity exponent of a crystalline two-dimensional membrane}

\author[S]{D. R. Saykin}
\author[KIT,TKM,PTI]{I. V. Gornyi}
\author[PTI]{V. Yu.~Kachorovskii}
\author[LITP,HSE]{I. S. Burmistrov\corref{cor}}

\address[S]{Department of Physics, Stanford University, Stanford, CA 94305, USA}
\address[KIT]{Institut f\"ur Nanotechnologie,  Karlsruhe Institute of Technology, 76021 Karlsruhe, Germany}
\address[TKM]{\mbox{Institut f\"ur Theorie der Kondensierten Materie, 76128 Karlsruhe, Germany}}
\address[PTI]{A. F.~Ioffe Physico-Technical Institute, 194021 St.~Petersburg, Russia}
\address[LITP]{L. D. Landau Institute for Theoretical Physics, Semenova 1-a, 142432, Chernogolovka, Russia}
\address[HSE]{Laboratory for Condensed Matter Physics, National Research University, Higher School of Economics, 101000 Moscow, Russia}
\cortext[cor]{Corresponding author. Fax: \texttt{+7-495-702-9317} }
\ead{burmi@itp.ac.ru}

\begin{abstract}
We compute the absolute Poisson's ratio $\nu$ and the bending rigidity exponent $\eta$ of a free-standing two-dimensional crystalline membrane embedded into a space of large dimensionality $d = 2 + d_c$, $d_c \gg 1$. 
We demonstrate that, in the regime of anomalous Hooke's law, the absolute Poisson's ratio approaches material independent value determined solely by the spatial dimensionality $d_c$: $\nu = -1 +2/d_c-a/d_c^2+\dots$ where $a\approx 1.76\pm 0.02$. Also, we find  the following expression for the exponent of the bending rigidity: $\eta = 2/d_c+(73-68\zeta(3))/(27 d_c^2)+\dots$. These results cannot be  captured by self-consistent screening approximation. 
\end{abstract}

\begin{keyword}
{crystalline membrane \sep tethered membrane \sep graphene \sep Poisson's ratio}
\end{keyword}

\end{frontmatter}

\section{Introduction}
\label{sec:intro}

The Mermin---Wagner theorem states that in two--dimensional (2D) crystals long--range order is destroyed due to thermal fluctuations \cite{MerminWagner,Hohenberg}. For $D=2$--dimensional membrane embedded into $d=3$--dimensional space, this means that transition from $O(d)$--symmetric crumpled phase to $O(D)\times O(d-D)$--symmetric flat phase governed by order parameter $m_{ab} = \partial_{b}r_{a}$ leads to existence of $d_c=d-D$  ``massless'' boson modes. Here $d$-dimensional vector $\bm{r}$ parametrizes the position of a point at the membrane, latin indices $a$ and $b$ indicate the spatial components of $\bm{r}$. Physically, this Goldstone boson  corresponds  to out-of-plane (or flexural) phonon mode $h(\bm{x})$, where 2D vector $\bm{x}=(x,y)$ parametrizes the surface of 2D membrane. The flexural phonon produces divergent contribution into thermal fluctuations of order parameter in the thermodynamic limit:
\begin{equation}
    \langle (\nabla {h})^2 \rangle
    \propto T \int \frac{d^2\bm{q}}{(2\pi)^2} \frac{q^2}{\varkappa q^4} 
    \propto \frac{T}{\varkappa}\ln \frac{L}{l}.
    \label{eq:MW}
\end{equation}
Here $\varkappa$ denotes the bending rigidity, $T$ is the temperature, $L$ stands for the size of the membrane, and $l$ is the ultra-violet cutoff of the order of the lattice spacing. In defiance to the statement of Eq. \eqref{eq:MW},  
free--standing 2D membrane, e.g. graphene, do exist experimentally. Resolution of the seeming paradox lies in the fact that Mermin---Wagner theorem applies only to the systems with short--range interactions.  Crystalline membranes posses long--range interaction between flexural phonons mediated by  in--plane phonons. Effectively such interaction leads to the stiffening of the membrane at large scales. In particular, at small momenta, $q\ll q_*$, the bending rigidity becomes renormalized \cite{AronovitzLubensky88}:
\begin{equation}
\varkappa_q = \varkappa (q_*/q)^\eta  .
\label{eq:bending}
\end{equation}
Here $q_* \sim \sqrt{T Y_0}/\varkappa$ is the so-called inverse Ginzburg length, where $Y_0 = \frac{4\mu(\mu+\lambda)}{2\mu+\lambda}$  denotes the Young modulus of the 2D crystalline membrane with $\mu$ and $\lambda$ being the Lam\'e coefficients of a material. The stiffening of the membrane
accounts for the existence of the flat phase:
\begin{equation}
    \langle(\nabla {h})^2\rangle 
    \propto \frac{T}{\varkappa}\int \frac{d^2\bm{q}}{(2\pi)^2} \frac{q^2}{\varkappa_q q^4}
    \propto \frac{T}{\eta \varkappa} .
\end{equation}
Thus, interaction between phonons is crucial for stability of the 2D membrane.  

Since harmonic approximation ($\eta=0$) does not suffice to be even a zeroth--order approximation and exact analytical solution of fully interacting problem of phonons modes is not feasible, one has to develop 
other methods.  Up to date, the exponent $\eta$ was determined within several approximate analytical schemes \cite{AronovitzLubensky88,paczuski1988,david1,david2,Doussal,Kownacki}. However, none of these approaches being controllable in the physical case $D=2$ and $d=3$.  Numerical simulations for the latter case yielded $\eta=0.60 \pm 0.10$ \cite{Gompper91}, $\eta=0.72 \pm 0.04$ \cite{Bowick96}, $\eta\approx 0.85$   \cite{Los-PRB-2009},  and $\eta=0.795\pm 0.01$ \cite{Troster}.

Non-trivial scaling of the bending rigidity, Eq. \eqref{eq:bending}, results in failure of the linear Hooke's law and in emergence of universal (i.e. material independent) Poisson's ratio in the regime of small tensions $\sigma\ll \sigma_*$ where $\sigma_* \simeq \varkappa q_*^2 \sim Y_0T/\varkappa$ \cite{buck,lower-cr-D2,david2,Doussal,nelson15,katsnelson16,my-hooke}. 
Recently, in the regime $\sigma \ll \sigma_*$ the anomalous Hooke's law, i.e. nonlinear dependence of the deformation on the stress, has been experimentally measured in graphene \cite{nicholl15}.

Most utilized analytical method to study the anomalous elastic properties of 2D crystalline membranes is the self--consistent screening approximation (SCSA) developed in seminal paper \cite{Doussal}. As any other self--consistent scheme, SCSA takes into account some subclass of diagrams in perturbation theory which is typically not preferable with respect to the others. This scheme becomes exact only in the limit $d_c\to \infty$ where it corresponds to the summation of the leading order logarithmic corrections in  perturbation theory. Within SCSA the following results for the bending rigidity exponent and the Posson's ratio (at zero external stress) have been obtained (see Ref. \cite{DoussalR} for a review): $\eta_{\rm SCSA} = 4/\bigl [d_c+\sqrt{16-2d_c+d_c^2}\bigr ]$ and $\nu_{\rm SCSA}=-1/3$. Surprisingly, for $d_c=1$ the value of $\eta_{\rm SCSA} \approx 0.82$ is very close to the result for $\eta$ reported from numerics. The value of the Poisson's ratio $\nu_{\rm SCSA}=-1/3$ is close to some numerical results for the Poisson's ratio in the physical case $d_c=1$ \cite{Bowick96,Falcioni1997}. The ``super-universal'' (i.e. independent of  $d_c$) SCSA result for the Poisson's ratio has been checked to be stable against inclusion of more diagrams in the self--consistent scheme  \cite{Gazit2009}. Also the ``super-universal'' SCSA result for the Poisson's ratio has been supported by the non-perturbative renormalization group treatment of the problem \cite{Hasselmann,Mouhanna}. 

The issue of the Poisson's ratio of  2D crystalline membrane occurs to be more complicated than it has been thought originally. In fact, due to the anomalous Hooke's law, the Poisson's ratio can be defined in many ways. Recently, two Poisson's ratios, differential and absolute, have been introduced and their behaviour has been studied \cite{PR-PRB}. The differential Poisson's ratio is determined as the ratio of change in displacements after application of the infinitesimally small uniaxial stress in addition to a finite isotropic tension. The absolute Poisson's ratio corresponds to the conventional definition of the Poisson's ratio, i.e. it is the ratio of displacements after application of a finite uniaxial stress. For $\sigma\ll \sigma_*$ both differential and absolute Poisson's ratios have universal but different values. In the limit $\sigma\to 0$ they coincide as follows from their definitions. However, their limiting value depends on the boundary conditions since the limit $\sigma\to 0$ and the membrane size $L\to \infty$ do not commute. Also, the differential and absolute Poisson's ratio coincide in the limit of linear Hooke's law, $\sigma\gg \sigma_*$ where they both are equal to the value $\lambda/(2\mu+\lambda)$ given by the classical elasticity theory \cite{LandauLifshitz}. The universal regime for the Poisson's ratios is realized at $\sigma_*\gg \sigma\gg \sigma_L$ where $\sigma_L = \sigma_* (q_* L)^{\eta-2}$. 

For some reasons, the corrections in $1/d_c$ to the results obtained within SCSA have not been analysed thoroughly. Recently, we have computed the differential Poisson's ratio to the first order in $1/d_c$ within the universal range of tensions, $\sigma_L\ll \sigma\ll \sigma_*$. We derived the following result: $\nu_{\rm diff} = -1/3+ c_{\rm diff}/d_c+\dots$ where the numerical constant $c_{\rm diff}\approx 0.016$ \cite{dPR}. 
This result indicates that the value of differential Poisson's ratio in the regime of the anomalous Hooke's law 
does depend on the number of flexural phonon modes, $d_c$. 

In this paper we extend analysis of $1/d_c$ corrections and compute them for the absolute Poisson's ratio $\nu$ and the bending rigidity exponent $\eta$. In particular, we find that 
\begin{equation}
\nu = -1 +\frac{2}{d_c}-\frac{a}{d_c^2}+\dots,
\label{eq:nu:main}
\end{equation}
where $a\approx 1.76\pm 0.02$ and
 \begin{equation}
 \eta = \frac{2}{d_c}-\frac{68\zeta(3)-73}{27 d_c^2}+\dots 
 \label{eq:eta:main}
 \end{equation}

The paper is organized as follows. In the section \ref{s1} a reader will find the description of the model we use to study 2D crystalline membrane and formal definitions of the absolute Poisson's ratio. In the Sec. \ref{s2} the perturbation theory in flexural phonon interaction is used to obtain expression for $\nu$ upto the seconf order in $1/d_c$. In the section \ref{s3} we calculate the critical exponent $\eta$ up to the second order in $1/d_c^2$. We end the paper with a summary of results, Sec. \ref{s4}. Technical details are given in Appendices.


\section{Formalism\label{s1}}

As our starting point we choose the effective action of the Landau--Ginzburg type introduced in the seminal papers \cite{AronovitzLubensky88,paczuski1988} for a free-standing 2D membrane. Its imaginary-time Lagrangian is written in terms of the $d$-dimensional vector $\bm{r}$:
\begin{gather}
\mathcal{L}[\bm{r}] =   \rho (\partial_\tau \bm{r})^2+ \frac{\varkappa}{2} (\triangle \bm{r})^2 +\frac{\mu}{4} \Bigl (\partial_\alpha \bm{r} \partial_\beta \bm{r}- \delta_{\alpha\beta}\Bigr )^2
+ \frac{\lambda}{8} \Bigl (\partial_\alpha \bm{r} \partial_\alpha \bm{r}- 2 \Bigr )^2  .
\label{eq:S:1}
\end{gather}
Here $\rho$ stands for the mass density of the membrane. The Greek indices correspond to the 2D coordinates $(x,y) \equiv \bm{x}$ parameterizing the membrane. 
To take into account the effect of external stress we introduce the stretching factors $\xi_x$ and $\xi_y$ such that:  $\bm{r} = \{\xi_x x+u_x,\xi_y y + u_y, h_1,\dots,h_{d_c}\}$. Here $\bm{u} = \{u_x,u_y\}$ corresponds to the in-plane phonons whereas $\bm{h}=\{h_1,\dots ,h_{d_c}\}$ describes the out-of-plane (or flexural) phonons. Substituting the reparametrization into Eq. \eqref{eq:S:1} allows one to write the partition function for a 2D crystalline membrane in terms of the following (see Refs. \cite{crump} for details):
\begin{equation}
Z = \int \mathcal{D}[\bm{u},\bm{h}]\, \exp (-S) .
\label{eq:PFZ}
\end{equation}
Here the action in the imaginary time is given by ($\beta=1/T$)
\begin{gather}
S = \int\limits_0^\beta d\tau \int d^2 \bm{x} \Biggl \{\frac{1}{8}\bigl ( 2\mu \delta_{\alpha\beta}+\lambda\bigr ) \Bigl [  \left (\xi_\alpha^2-1+{K}_\alpha\right )\left (\xi_\beta^2-1+{K}_\beta\right ) - {K}_\alpha {K}_\beta  \Bigr ]
 +\frac{\rho}{2} \Bigl [ (\partial_\tau \bm{u})^2 +(\partial_\tau \bm{h})^2\Bigr ] \notag \\
 +
\frac{\varkappa}{2}\Bigl [ (\Delta \bm{h})^2+(\Delta \bm{u})^2\Bigr ] + \mu {u}_{\alpha\beta}  {u}_{\beta\alpha}
+\frac{\lambda}{2} {u}_{\alpha\alpha} {u}_{\beta
\beta} \Biggr \} ,
\label{eq:action:i}
\end{gather}
where
\begin{equation}
{u}_{\alpha\beta} = \frac{1}{2} \Bigl ( \xi_\beta \partial_\alpha u_\beta +\xi_\alpha \partial_\beta u_\alpha + \partial_\alpha \bm{u}  \partial_\beta \bm{u} + \partial_\alpha \bm{h} \partial_\beta \bm{h} \Bigr ) ,
\label{eq:uab:def}
\end{equation}
and
\begin{equation}
{K}_\alpha =\frac{1}{\beta L^2} \int\limits_0^\beta d\tau \int d^2 \bm{x}\, \Bigl (\partial_\alpha \bm{u}  \partial_\alpha \bm{u} + \partial_\alpha \bm{h} \partial_\alpha \bm{h}\Bigr )  .
\label{eq:K:def}
\end{equation}

In this paper we limit the analysis of the absolute Poisson's ratio to the case of low enough temperature, $T\ll \varkappa$ \cite{crump}. 
This condition allows us to neglect the term $\partial_\alpha \bm{u} \partial_\beta \bm{u}$ in comparison with  $\partial_\alpha \bm{h} \partial_\beta \bm{h}$ (see the expressions for ${u}_{\alpha\beta}$ and $K_\alpha$ in Eqs. \eqref{eq:uab:def} and \eqref{eq:K:def}, respectively) \cite{footnoteCrumpling}. Then we can simplify the effective action \eqref{eq:action:i} by 
integrating the in-plane phonons \cite{Peliti1987} such that the partition function becomes an integral over static flexural phonons only (see Ref. \cite{crump} for details):
\begin{equation}
Z = \int \mathcal{D}[\bm{h}]\, \exp (-E/T) .
\end{equation}
Here the energy $E$ of a given configuration of the flexural phonon field $h(\bm{x})$ is as follows 
\begin{gather}
E =   \frac{L^2}{8}\left (\xi_\alpha^2-1+{\tilde K_\alpha}\right ) M_{\alpha\beta} \left (\xi_\beta^2-1+{\tilde K_\beta}\right ) 
  + \frac{\mu}{2 L^2}\left ( \int d^2 \bm{x}  \
  \partial_x \bm{h} \partial_y \bm{h}\right )^2+\frac{\varkappa}{2} \int d^2 \bm{x}\  (\Delta \bm{h})^2
   \notag \\
 + \frac{Y_0}{8} \int^\prime \frac{d^2 \bm{k} d^2\bm{k^\prime} d^2 {\bm q}}{(2\pi)^6}
  \frac{[\bm{k}\times \bm{q}]^2}{q^2}\frac{[\bm{k^\prime}\times \bm{q}]^2}{q^2} \bigl (\bm{h}_{\bm{k}+\bm{q}} \bm{h}_{-\bm{k}}\bigr )\bigl (
  \bm{h}_{-\bm{k^\prime}-\bm{q}} \bm{h}_{\bm{k^\prime}}\bigr ) .
\label{eq:action:iv}
\end{gather}
Here we introduced the $2\times 2$ matrix $M_{\alpha\beta} = 2\mu \delta_{\alpha\beta}+\lambda$. The quantity ${\tilde K}_\alpha$ is obtained from ${K}_\alpha$ by omitting the term $\partial_\alpha \bm{u} \partial_\alpha \bm{u}$:
\begin{equation}
{\tilde K}_\alpha =\frac{1}{L^2} \int d^2 \bm{x}\,  \partial_\alpha \bm{h} \partial_\alpha \bm{h}  .
\label{eq:tildeK:def}
\end{equation}
The `prime' sign in the last integral on the right hand side of Eq. \eqref{eq:action:iv} indicates that the interaction of the flexural phonons with $q=0$ is excluded. 

The partition function $Z$ depends on the stretching factors $\xi_x$ and $\xi_y$. The  diagonal components of the tension tensor are determined as 
\begin{equation}
\sigma_x = \frac{1}{\xi_x} \frac{\partial f}{\partial \xi_x}, \qquad
\sigma_y = \frac{1}{\xi_y} \frac{\partial f}{\partial \xi_y} ,
\label{eq:EOS}
\end{equation}
where $f = -T L^{-2}\ln Z$ denotes the free energy per unit area. We note that Eq. \eqref{eq:EOS} 
is the equation of state which determines the relation between the tension tensor $\{\sigma_x, \sigma_y\}$ and the stretching tensor $\{\xi_x,\xi_y\}$.

In what follows, it will be more convenient to choose the diagonal components of the tension tensor as independent variables rather than $\xi_x$ and $\xi_y$. As usual, the corresponding free energy $g(\sigma_x,\sigma_y)$ can be constructed from $f(\xi_x,\xi_y)$ via the Legendre transform:
\begin{equation}
g(\sigma_x,\sigma_y) = f(\xi_x,\xi_y) - \sigma_x (\xi^2_x-1)/2-\sigma_y(\xi_y^2-1)/2 ,
\end{equation}
where $\xi_\alpha$ is expressed in terms of $\sigma_\alpha$ with the help of the equation of state \eqref{eq:EOS}. In the thermodynamic limit, $L\to \infty$, one can explicitly write that 
\begin{equation}
g = - T L^{-2} \ln \mathcal{Z}, \qquad \mathcal{Z} = \int \mathcal{D}[\bm{h}]\, \exp (-\mathcal{E}/T) ,
\end{equation}
where 
\begin{gather}
\mathcal{E} =   -\frac{L^2}{2} \sigma_\alpha M^{-1}_{\alpha\beta} \sigma_\beta 
  + \frac{\mu}{2 L^2}\left ( \int d^2 \bm{x}  \
  \partial_x \bm{h} \partial_y \bm{h}\right )^2+\frac{\varkappa}{2} \int d^2 \bm{x}\  (\Delta \bm{h})^2
+   \frac{\sigma_\alpha}{2} \int d^2 \bm{x}\  (\nabla_\alpha \bm{h})^2
   \notag \\
 + \frac{Y_0}{8} \int^\prime \frac{d^2 \bm{k} d^2\bm{k^\prime} d^2 {\bm q}}{(2\pi)^6}
  \frac{[\bm{k}\times \bm{q}]^2}{q^2}\frac{[\bm{k^\prime}\times \bm{q}]^2}{q^2} \bigl (\bm{h}_{\bm{k}+\bm{q}} \bm{h}_{-\bm{k}}\bigr )\bigl (
  \bm{h}_{-\bm{k^\prime}-\bm{q}} \bm{h}_{\bm{k^\prime}}\bigr ) .
\label{eq:action:vg}
\end{gather}
In terms of $g(\sigma_{x},\sigma_{y})$ the equation of states can be written as
\begin{equation}
\varepsilon_x\equiv \frac{\xi_x^2-1}{2} = - \frac{\partial g}{\partial \sigma_x}, \qquad \varepsilon_y\equiv \frac{\xi_y^2-1}{2} = - \frac{\partial g}{\partial \sigma_y} ,
\end{equation} 
or more explicitly,
\begin{gather}
\begin{pmatrix}
\sigma_x\\
\sigma_y
\end{pmatrix}
= M
\begin{pmatrix}
\varepsilon_x+\langle \tilde K_x \rangle/2 \\
\varepsilon_y+\langle \tilde K_y\rangle/2
\end{pmatrix} .
\label{eq:EOS:2}
\end{gather}
Here the average $\langle\dots \rangle$ is with respect to
the energy \eqref{eq:action:vg}. 

Now the absolute Poisson's ratio can be defined as follows. Let us apply the uniaxial stress, $\sigma_{x}=\sigma$ and $\sigma_{y}=0$, and consider the change of the stretching factors:
\begin{equation}
\delta \varepsilon_\beta(\sigma) = \varepsilon_\beta(\sigma,0)- \varepsilon_\beta(0,0) .
\end{equation} 
Then the absolute Poisson's ratio is given as 
\begin{equation}
\nu =- \frac{\delta \varepsilon_y}{\delta \varepsilon_x} =
\frac{\nu_0 - Y_0 \delta \tilde{K}_y/(2\sigma)}{1 + Y_0\delta \tilde{K}_x/(2\sigma)},
\label{eq:nu:def:1}
\end{equation} 
where $\delta \tilde{K}_\beta(\sigma) = \langle\tilde{K}_\beta(0,0)\rangle-\langle\tilde{K}_\beta(\sigma,0)\rangle$ and $\nu_0=\lambda/(2\mu+\lambda)$ stands for the classical value of the Poisson's ratio. 

The overall behavior of the absolute Poisson ratio on $\sigma$ has been discussed recently in Ref. \cite{PR-PRB}. Since $Y_0\delta \tilde{K}_\beta(\sigma)/\sigma \sim (\sigma_*/\sigma)^{1-\bm{\alpha}}$ with the exponent $\bm{\alpha} = \eta/(2-\eta)$ \cite{lower-cr-D2}, the general expression
\eqref{eq:nu:def:1} for the absolute Poisson's ratio can be simplified in the universal regime $\sigma\ll \sigma_*$:
\begin{equation}
\nu =- \frac{\delta \tilde{K}_y}{\delta \tilde{K}_x} .
\label{eq:nu:def:2}
\end{equation}


\section{Perturbation theory in $1/d_c$ for the absolute Poisson's ratio\label{s2}}

In order to proceed with the computation of the absolute Poisson's ratio, Eq. \eqref{eq:nu:def:2}, we need to compute the $\delta\tilde{K}_\alpha(\sigma)$. We can write the following formal expression in terms of the exact Green's function of flexural phonons:
\begin{equation}
\langle\tilde{K}_\beta(\sigma,0) \rangle=
d_c \int \frac{d^2\bm{q}}{(2\pi)^2} q_\beta^2
\mathcal{G}_q, \qquad 
\mathcal{G}_q = \frac{T}
{\varkappa q^4+\sigma q_x^2-\Sigma_\sigma(\bm{q})} .
\end{equation}
As usual, the self-energy $\Sigma_\sigma(\bm{q})$ is due to the interaction between out-of-plane phonons.
Although at $\sigma\ll \sigma_*$ this interaction is effectively controls by $1/d_c$ we cannot develop the expansion in $1/d_c$ by expanding of the Green's function in powers of $\Sigma_\sigma(\bm{q})$. The point is the infra-red divergence of the expression for $K_\beta(\sigma,0)$ in the absence of $\Sigma_\sigma(\bm{q})$. In order to resolve this problem we construct an expansion in difference 
$\delta \Sigma_\sigma(\bm{q})  = \Sigma_\sigma(\bm{q}) - \Sigma_0(q)$. Then, we find 
\begin{equation}
\delta \tilde{K}_\beta(\sigma) = 
d_c T \int \frac{d^2\bm{q}}{(2\pi)^2} \frac{q_\beta^2[\sigma q_x^2-\delta \Sigma_\sigma(\bm{q})]}{\varkappa_q q^4 [\varkappa_q q^4+\sigma q_x^2 -\delta \Sigma_\sigma(\bm{q})]} ,
\end{equation}
where $\varkappa_q q^4 = \varkappa q^4 - \Sigma_0(q)$. Now we can expand the self-energy  difference $\delta \Sigma_\sigma(\bm{q})$ in powers of $1/d_c$: 
\begin{equation}
\delta \Sigma_\sigma(\bm{q}) =
\delta \Sigma_\sigma^{(1)}(\bm{q}) + \delta \Sigma_\sigma^{(2)}(\bm{q}) + \dots
\end{equation}
Such an expansion results in regular perturbation series for $\delta \tilde{K}_\beta(\sigma)$:
\begin{equation}
\delta \tilde{K}_\beta(\sigma) =
\delta \tilde{K}^{(0)}_\beta(\sigma) +\delta \tilde{K}^{(1)}_\beta(\sigma) + \dots
\end{equation}
Then, we find
\begin{equation}
\nu = - \frac{\delta \tilde{K}^{(0)}_y(\sigma)}{\delta \tilde{K}^{(0)}_x(\sigma)}
\left ( 
1 + \frac{\delta \tilde{K}^{(1)}_y(\sigma)}{\delta \tilde{K}^{(0)}_y(\sigma)} - \frac{\delta \tilde{K}^{(1)}_x(\sigma)}{\delta \tilde{K}^{(0)}_x(\sigma)}
+ \dots
\right )  .
\end{equation}
Here the functions $\delta \tilde{K}^{(0,1)}_\beta(\sigma)$ are given explicitly as
\begin{equation}
\delta \tilde{K}^{(0)}_\beta(\sigma)
= d_c T\int \frac{d^2\bm{q}}{(2\pi)^2} \frac{\sigma  q_\beta^2 q_x^2}{\varkappa_q q^4 [\varkappa_q q^4+\sigma q_x^2 ]}
\end{equation}
and
\begin{equation}
\delta \tilde{K}^{(1)}_\beta(\sigma)
= - d_c T\int \frac{d^2\bm{q}}{(2\pi)^2} \frac{q_\beta^2 \delta \Sigma_\sigma^{(1)}(\bm{q})}{[\varkappa_q q^4+\sigma q_x^2 ]^2} .
\end{equation}
Both expressions are manifestly convergent in the ultra-violet. For the convergence of $\delta \tilde{K}^{(0)}_\beta(\sigma)$ in the infrared, the power-law renormalization of the bending rigidity (due to the interaction-induced self-energy $\Sigma_0(\bm{q})$)  is crucial. The expression  $\delta \tilde{K}^{(1)}_\beta(\sigma)$ is convergent in the infrared even in the absence of the bending rigidity renormalization.

Performing evaluation of the integral over momentum we find
\begin{equation}
\delta \tilde{K}^{(0)}_\beta(\sigma) = 
\frac{d_c T}{2 (2-\eta) \varkappa_*}\left(\frac{\sigma}{\sigma_*}\right )^\alpha
\left (\sin \frac{\pi\eta}{2-\eta}\right )^{-1}
\langle n_x^{2\bm{\alpha}} n_\beta^2\rangle  .
\end{equation}
We note that since $\eta=2/d_c+O(1/d_c^2)$  we obtain the following result at $d_c\to \infty$ ($\eta \to 0$)
\begin{equation}
\delta \tilde{K}^{(0)}_\beta(\sigma) = \frac{d_c^2 T}{8\pi \varkappa} .
\end{equation}
Now using the relation 
\begin{equation}
\frac{\langle n_x^{2\bm{\alpha}} n_y^2\rangle}{\langle n_x^{2\bm{\alpha}} n_x^2\rangle} = \frac{1}{1+2\bm{\alpha}} ,
\end{equation}
we find
\begin{equation}
\nu =  - \frac{1}{1+2\bm{\alpha}} \left ( 
1-\frac{c}{d_c^2}
+\dots
\right ), \qquad 
c = \frac{8\pi \varkappa}{T}
\Bigl [\delta \tilde{K}^{(1)}_x(\sigma)-
\delta \tilde{K}^{(1)}_y(\sigma)\Bigr ] .
\label{eq:c:1}
\end{equation}
Since the integrals in $\delta \tilde{K}^{(1)}_\beta(\sigma)$ are convergent even in the absence of the renormalization of the bending rigidity, we can set $\eta=0$ in $\varkappa_q$ for the computation of the constant $c$. Taking into account the deviation of $\eta$ from $0$ results in the $1/d_c$ correction to $c$ which is beyond our accuracy.

\begin{figure}
\centerline{\includegraphics[width=0.25\textwidth]{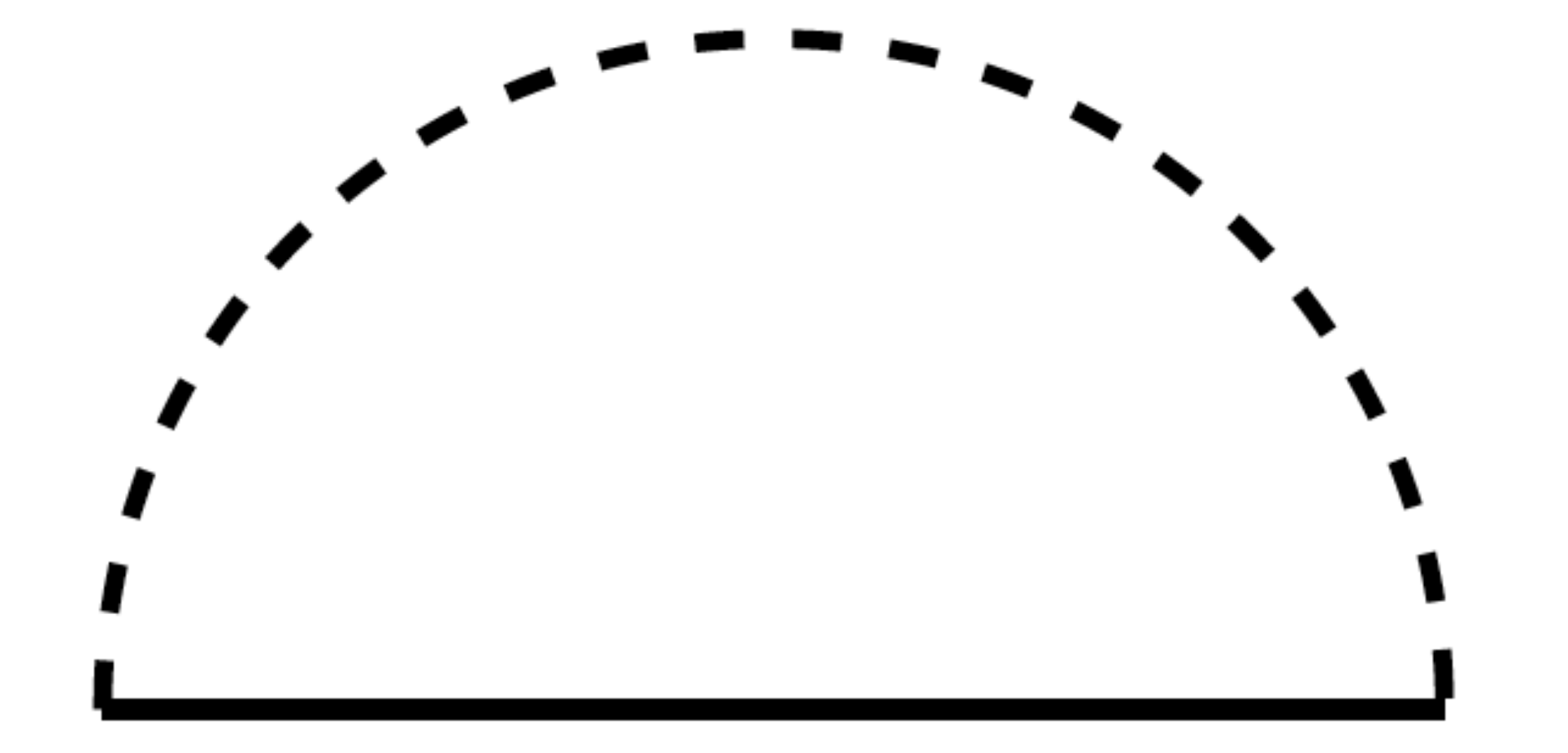}}
\caption{The self-energy correction of the first order in the screened interaction.}
\label{Fig1}
\end{figure}

\subsection{Evaluation of $\delta K^{(1)}_\beta(\sigma)$}

In order to find the value of the coefficient $c$, we need to compute $\delta K^{(1)}_\beta(\sigma)$. The self-energy $\Sigma^{(1)}_\sigma({\bm{k}})$ is shown in Fig. \ref{Fig1}. 
It involves the screened interaction \cite{Doussal} between flexural phonons (for details see Appendix A of Ref. \cite{dPR}): 
\begin{equation}
N^{(\sigma)}_q = \frac{Y_0/2}{1+ 3 Y_0 \Pi^{(\sigma)}_q/2} .
\label{eq:Nprime}
\end{equation}
Here $\Pi_q^{(\sigma)}$ denotes the irreducible polarization operator in the presence of uniaxial stress. To the leading order in $1/d_c$ it can be written as
\begin{equation}
\Pi^{(\sigma)}(\bm{q}) = \frac{d_c T}{3} \int \frac{d^2 \bm{k}}{(2\pi)^2} \frac{[\bm{k}\times\bm{q}]^4}{q^4} \frac{1}{[\varkappa k^4+\sigma k_x^2][\varkappa |\bm{k}-\bm{q}|^4+\sigma (k_x-q_x)^2]} .
\label{eq:Pi:def:0}
\end{equation}
The polarization operator has the following scaling form:
\begin{equation}
\Pi^{(\sigma)}(\bm{q}) = \frac{d_c T}{3\varkappa \sigma}
\mathcal{P}(\bm{q}/q_\sigma) .
\end{equation}
In the universal regime $\sigma\ll \sigma_*$ one can neglect unity in denominator of Eq. \eqref{eq:Nprime} and approximate  $N^\prime_q $ as $1/[3 \Pi^{(\sigma)}_q]$. Then the corresponding self-energy difference becomes 
\begin{equation}
\delta \Sigma^{(1)}_\sigma({\bm{k}}) = 
\frac{2 T}{3} \int \frac{d^2\bm{q}}{(2\pi)^2} \frac{[\bm{k}\times\bm{q}]^4}{q^4} 
\left (
\frac{1}{\varkappa |\bm{k}-\bm{q}|^4}\frac{1}{\Pi_q^{(0)}}
- \frac{1}{\varkappa |\bm{k}-\bm{q}|^4+\sigma |\bm{k}-\bm{q}|^2}\frac{1}{\Pi_q^{(\sigma)}} \right ) .
\label{eq:Sigma1}
\end{equation}

Surprisingly, the explicit expression for the function $\mathcal{P}(\bm{Q})$ can be found analytically (see   
 \ref{App1} for details). Introducing the function $A(Q) = \arcsinh (Q)/[Q\sqrt{1+Q^2}]$ and the vector $\bm{P}=(Q_x+i, Q_y)$, we result can be written as
 \begin{gather}
\mathcal{P}(\bm{Q}) =\Biggl \{ (1-4\hat{Q}_x^2)\hat{Q}_y^2\Bigl\{ \Bigl (A(Q) - 1\Bigr )
             - \re\Bigl [Q^2P^{-2} \bigl (A(P) - 1\bigr ) \Bigr ]\Bigr\} 
             + (1+Q^2)A(Q)+\re \Bigl[ (1-P^2)A(P) \Bigr]
     \notag \\
     + 4\hat{Q}_x\hat{Q}_y^2\bigl (\hat{Q}_y^2-\hat{Q}_x^2\bigr )
        \im\Bigl [QP^{-2} \bigl (A(P)-1\bigr ) \Bigr ] 
         + 4\hat{Q}_x^2\hat{Q}_y^2 \Bigl [A(Q)-\re A(P)\Bigr ]
     + 2(\hat{Q}_x^2-\hat{Q}_y^2)\im \Bigl [ P_x A(P) \Bigr ] \notag\\ 
     + 4\hat{Q}_x\hat{Q}_y \im \Bigl[ P_y A(P)\Bigr ] \Biggr \}/(8\pi) ,
     \label{eq:Pi:final}
\end{gather}
\color{black}
where $\bm{\hat Q} = \bm{Q}/Q = \{\cos\phi,\sin\phi\}$ and $P=\sqrt{(Q_x+i)^2+Q_y^2}$. The function $\mathcal{P}(\bm{Q})$ has the following asymptotic behaviour:
\begin{gather}
\mathcal{P}(\bm{Q}) = \begin{cases}
\displaystyle \frac{\cos^{7/2}\phi}{8\sqrt{Q}}-\frac{2\cos(2\phi)+\cos(4\phi)}{8\pi}, \quad Q\ll 1 \\
\displaystyle \frac{3}{16 \pi Q^2} +
\frac{9 [1 + 4 \ln (2 Q)] - 
 8 \cos(2\phi) [4 - 3 \ln(2Q)]+\cos(4 \phi) }{192 \pi Q^4}
,  \quad Q\gg 1
\end{cases}
\end{gather}
We note the strong anisotropy in angle dependence of $\mathcal{P}(\bm{Q})$ at small $Q$. 
The asymptotics of $\mathcal{P}(\bm{Q})$ at large $Q$ implies that 
\begin{equation}
\Pi^{(0)}(\bm{q}) = \frac{d_c T}{16\pi \varkappa^2 q^2} .
\label{eq:Pi0}
\end{equation}
Now using Eqs. \eqref{eq:Sigma1}, \eqref{eq:Pi:final}, and \eqref{eq:Pi0}, we can rewrite Eq. \eqref{eq:c:1} in the following form
\begin{gather}
c= 16\pi \int \frac{d^2\bm{Q}d^2\bm{K}}{(2\pi)^4}\frac{[\bm{K}\times\bm{Q}]^4}{Q^4}(K_x^2-K_y^2)
\Biggl[ 
 \frac{1}{\mathcal{P}(\bm{Q})[K^4+K_x^2]^2[|\bm{K}-\bm{Q}|^4+(K_x-Q_x)^2]}
 \notag \\
 - 
 \frac{1}{\mathcal{P}_0(Q)[K^4+K_x^2]^2 |\bm{K}-\bm{Q}|^4}
\Biggr ] ,
\label{eq:express:c}
\end{gather}
where $\mathcal{P}_0(Q) = 3/(16\pi Q^2)$. Numerical evaluation of this integral (see \ref{App2}) yields
\begin{equation}
c = 0.56 \pm 0.02  .
\label{eq:c:final}
\end{equation}

Now using Eq. \eqref{eq:c:1}, we can write the expansion of the absolute Poisson ratio to the second order in $1/d_c^2$:
\begin{equation}
\nu =   -
1+\frac{2}{d_c} -
\frac{2-b-c}{d_c^2}
+\dots .
\label{eq:nu:inter}
\end{equation}
Here the coefficient $b$ determines expansion of the bending rigidity exponent to the second order in $1/d_c^2$:
\begin{equation}
\eta = \frac{2}{d_c}+\frac{b}{d_c^2} + \dots .
\end{equation}
Therefore, in order to determine the absolute Poisson ratio to the second order in $1/d_c$ one needs to 
compute $\eta$ to the same order.

\section{Evaluation of $\eta$ to the second order in $1/d_c$\label{s3}}

Perturbative calculation of critical exponent $\eta$ describing softening of the flexural mode due to interaction between phonons is quite straightforward. General statement \eqref{eq:bending} for $\varkappa_q$ implies that exact self-energy $\Sigma_0(k)$ has the following expansion at small values of momenta, $k\ll q_*$, and for $\eta \ll 1$:
\begin{gather}
\Sigma_0(k) = - \varkappa k^4 \Bigl ( \eta \ln\frac{q_*}{k} + \frac{\eta^2}{2}\ln^2\frac{q_*}{k} + \dots
\Bigr ) .
\label{eq:Sigma0:exp}
\end{gather}

\subsection{SCSA type contributions to $\eta$}

\subsubsection{First order in $1/d_c$ correction to the self-energy}

In order to set notations, we start from the self-energy correction in the first order in the screened interaction 
(see Fig. \ref{Fig1}): 
\begin{equation}
\Sigma^{(1)}_0(k) = - 2 \int_q \frac{[\bm{k \times q}]^4}{q^4}
N_q^{(0)} \mathcal{G}^{(0)}_{\bm{k-q}} , \qquad \mathcal{G}^{(0)}_k= \frac{T}{\varkappa k^4} .
\end{equation}
Here we introduced for a brevity the following shorthand notation: $\int_q \equiv \int d^2\bm{q}/(2\pi)^2$.
Let us define $\tilde{q}_* = \sqrt{3 d_cY_0 T/(32 \pi \varkappa^2)}$. Then, we obtain
\begin{equation}
\Sigma^{(1)}_0(k) = - \frac{2}{d_c} \frac{16\pi \varkappa}{3} \int_q \frac{[\bm{k \times q}]^4}{q^2}
\frac{\tilde{q}_*^2}{\tilde{q}_*^2+q^2} \frac{1}{|\bm{k-q}|^4} ,
\end{equation}
Using the following integral
\begin{equation}
\int_0^{2\pi} \frac{d\theta}{2\pi} 
\frac{\sin^4\theta}{(1+q^2-2 q\cos\theta)^2} =
\frac{3}{8} \begin{cases}
1, & \quad q\leqslant 1, \\
q^{-4}, & \quad q>1 ,
\end{cases}
\end{equation}
we obtain
\begin{equation}
\Sigma^{(1)}_0(k) = - \frac{2}{d_c} \varkappa k^4  L(k/\tilde{q}_*), 
\end{equation}
where 
\begin{gather}
L(K)=\frac{1}{K^4} \int_0^K dq  \frac{q^3}{1+q^2} + \int_K^\infty dq  \frac{1}{q(1+q^2)} =
-\ln K + \frac{1}{2}\ln(1+K^2) + \frac{K^2-\ln(1+K^2)}{2K^4}.
\end{gather}
For $k/\tilde{q}_* \ll 1$ we find
\begin{equation}
\Sigma^{(1)}_0(k) = - \frac{2}{d_c}\varkappa k^4 \ln \frac{q_*}{k}
, 
\label{eq:Sigma:1}
\end{equation}
where $q_* = e^{1/4} \tilde{q}_*$. Comparison of this result with the expansion \eqref{eq:Sigma0:exp} yields 
$\eta= 2/d_c$. 

\begin{figure}
\centerline{(a) \includegraphics[width=0.25\textwidth]{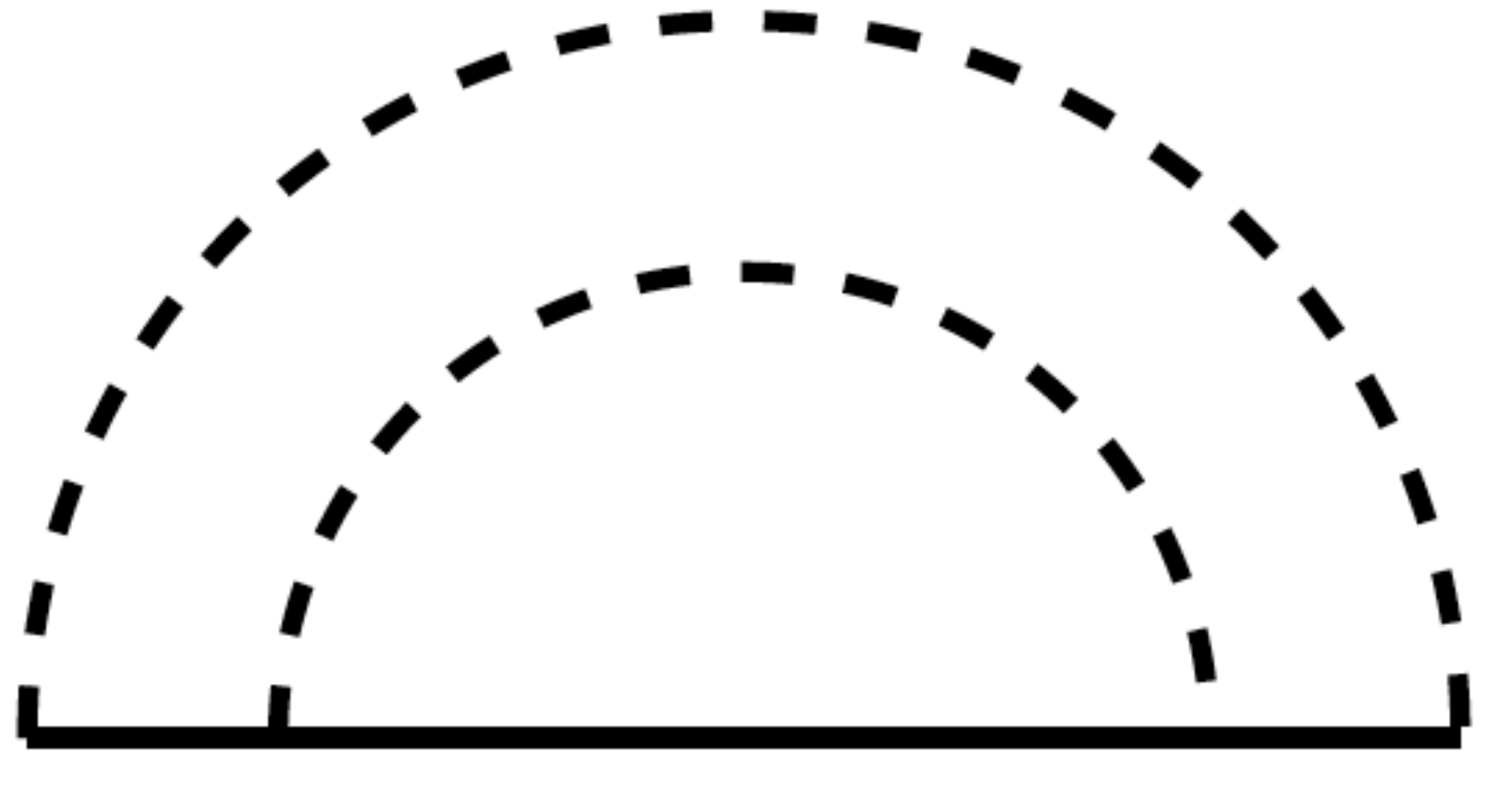}\qquad\qquad
(b) \includegraphics[width=0.25\textwidth]{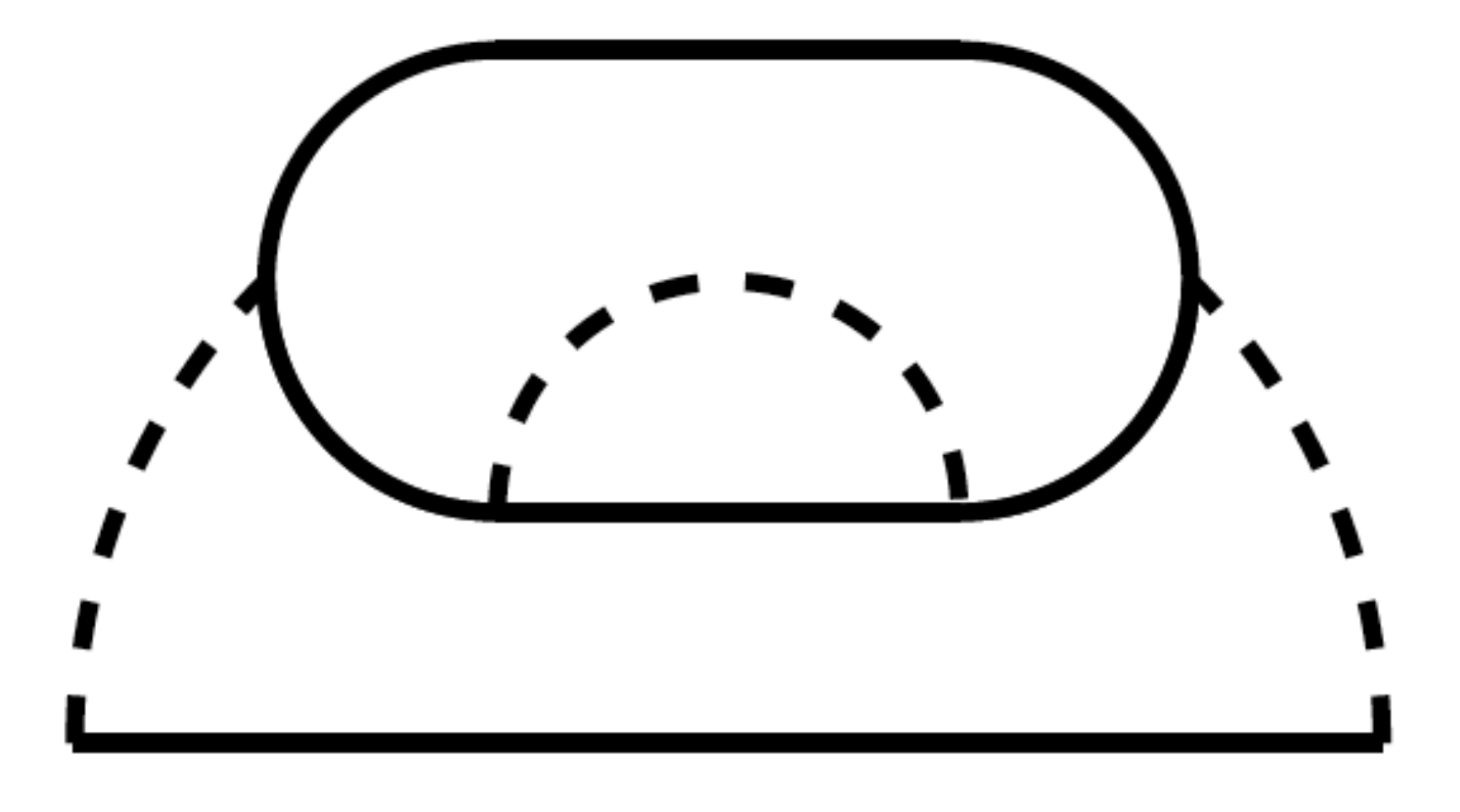}}
\caption{SCSA-type self-energy correction in the second order in $1/d_c$.}
\label{Fig2}
\end{figure}

\subsubsection{Second order self-energy correction}

We start from the diagram (a) in Fig. \ref{Fig2}. The corresponding contribution to the self-energy can be written as
\begin{equation}
\Sigma^{(2,a)}_0(k) = - \frac{2}{T} \int_q \frac{[\bm{k \times q}]^4}{q^4}
N_q^{(0)} \Bigl [\mathcal{G}^{(0)}_{\bm{k-q}}\Bigr ]^2\Sigma^{(1)}_0(|\bm{k-q}|)
= \frac{4}{d_c^2} \frac{16\pi\varkappa}{3} 
\int_q \frac{[\bm{k \times q}]^4 \tilde{q}_*^2}{q^2(\tilde{q}_*^2+q^2)} \frac{L(|\bm{k-q}|/\tilde{q}_*) }{|\bm{k-q}|^4} .
\end{equation}
This diagram diverges in the infrared as $\ln^2(k/\tilde{q}_*)$. However, we are interested also in the next, subleading, term which behaves as $\ln(k/\tilde{q}_*)$. Therefore, we cannot approximate the function $L(|\bm{k-q}|/\tilde{q}_*)$ by the logarithm. Instead, we rewrite $\Sigma^{(2,a)}_0(k)$ as follows 
\begin{gather}
\Sigma^{(2,a)}_0(k) = \frac{4}{d_c^2} \frac{16\pi\varkappa}{3} 
\int_q \frac{[\bm{k \times q}]^4}{q^2}\frac{\tilde{q}_*^2}{\tilde{q}_*^2+q^2} \frac{L(q/\tilde{q}_*)}{|\bm{k-q}|^4} 
+ \frac{4}{d_c^2} \frac{16\pi\varkappa}{3} 
\int_q \frac{[\bm{k \times q}]^4}{q^2}\frac{\tilde{q}_*^2}{\tilde{q}_*^2+q^2} \frac{1}{|\bm{k-q}|^4} 
\notag \\
\times 
\Bigl [ L(|\bm{k-q}|/\tilde{q}_*) - L(q/\tilde{q}_*)\Bigr ] .
\end{gather}
The last integral in the right hand side of the above expression is convergent in both ultraviolet and infrared. Thus we are not interested in it. Then, we find
\begin{gather} 
\Sigma^{(2,a)}_0 (k) = \frac{4}{d_c^2} \varkappa k^4
\left [ \frac{1}{K^4} \int_0^K dq \frac{q^3 L(q)}{1+q^2}  + \int_K^\infty dq \frac{L(q)}{q(1+q^2)} \right ] 
\end{gather}
Evaluating the integrals for $K\ll 1$, we obtain
\begin{gather}
\Sigma^{(2,a)}_0 (k) =
\frac{2}{d_c^2}\varkappa k^4 \Bigl [\ln^2K-\ln K\Bigr ]
= \frac{2}{d_c^2}\varkappa k^4 \Biggl [\ln^2(q_*/k)+\frac{1}{2} \ln (q_*/k)\Biggr ]
\label{eq:Sigma:2a}
\end{gather}

Next, we compute the diagram (b) in Fig. \ref{Fig2}. The diagram can be considered as the first order correction to the self-energy in which the interaction line is changed due to correction to the polarization operator: 
\begin{equation}
\Sigma^{(2,b)}_0(k) = \frac{6}{T} \int_q \frac{[\bm{k \times q}]^4}{q^4}
\Bigl [N^{(0)}_q\Bigr ]^2 \mathcal{G}^{(0)}_{\bm{k-q}}\delta \Pi^{(0)}_q
, \qquad \delta \Pi^{(0)}_q = \frac{2d_c}{3T}
\int_k \frac{[\bm{k \times q}]^4}{q^4}
\mathcal{G}^{(0)}_{\bm{k-q}}\Bigl [ \mathcal{G}^{(0)}_k\Bigr] ^2
\Sigma^{(1)}_{0}(k) .
\end{equation}
The correction to the polarization operator becomes
\begin{gather}
\delta \Pi^{(0)}_q = 
- \frac{4T^2}{3\varkappa^2}
\int_k \frac{[\bm{k \times q}]^4}{q^4} \frac{L(k/\tilde{q}_*)}{k^4|\bm{k-q}|^4}
 =
- \frac{T^2}{4\pi \varkappa^2q^2}
\Bigl [ \int_0^1 dk k L(q k/\tilde{q}_*)+
\int_1^\infty \frac{dk}{k^3}  L(q k/\tilde{q}_*) \Bigr ] 
\notag \\
=
- \frac{T^2}{4\pi \varkappa^2q^2}\tilde{L}(q/\tilde{q}_*) ,
\end{gather}
where  
\begin{equation}
\tilde{L}(K) =
- \frac{3+K^2}{3} \ln K 
+\frac{(1+K^2)}{6K^2}\Bigl[ \frac{(1+K^2)^2}{K^2}\ln(1+K^2)-1\Bigr ]
\end{equation}
We note that the function $\tilde{L}(K)$ has the same asymptotic behaviour at $K\ll 1$ as the function ${L}(K)$. At $K\gg 1$ the asymptotic of  $\tilde{L}(K)$ is given as $1/(2K^2)$. Then, we obtain
\begin{gather}
\Sigma^{(2,b)}_0(k) = 
- \frac{8}{d_c^2} \frac{16\pi \varkappa}{3} \int_q 
\frac{[\bm{k \times q}]^4}{q^2|\bm{k-q}|^4}
\frac{\tilde{q}_*^4 \tilde{L}(q/\tilde{q}_*)}{(\tilde{q}_*^2+q^2)^2} 
 = - \frac{8}{d_c^2} \varkappa k^4 \Biggl [
\frac{1}{K^4}\int_0^K dq \frac{q^3 \tilde{L}(q)}{(1+q^2)^2} 
\notag \\ + \int_K^\infty dq \frac{\tilde{L}(q)}{q(1+q^2)^2} \Biggr ] .
\end{gather}
At $K\ll 1$ we find
\begin{gather}
\Sigma^{(2,b)}_0(k) = 
- \frac{4}{d_c^2}  \varkappa k^4 \Bigl [ \ln^2 K-\ln K\Bigr ]
= - \frac{4}{d_c^2}  \varkappa k^4 \Biggl[ \ln^2({q}_*/k)+\frac{1}{2} \ln ({q}_*/k)\Biggr ] .
\label{eq:Sigma:2b}
\end{gather}

Summing up the SCSA-type corrections to the self-energy, \eqref{eq:Sigma:1}, \eqref{eq:Sigma:2a}, and \eqref{eq:Sigma:2b}, we find that 
\begin{gather}
\varkappa k^4- \Sigma^{(1)}_0(k) - \Sigma^{(2,a)}_0(k)-\Sigma^{(2,b)}_0(k) = \varkappa k^4 \Bigl [ 1 +\frac{2}{d_c} \ln \frac{q_*}{k} +\frac{2}{d_c^2} \ln^2 \frac{q_*}{k}+ \frac{1}{d_c^2}\ln \frac{q_*}{k}\Bigr ]  \approx \varkappa k^4 \left (\frac{q_*}{k}\right )^{\eta_{\rm SCSA}},
\label{eq:Sigma:SCSA}
\end{gather} 
where $\eta_{SCSA} = {2}/{d_c}+{1}/{d_c^2}+\dots$ which is nothing but expansion of the general SCSA result in $1/d_c$.

\begin{figure}
\centerline{(c) \includegraphics[width=0.25\textwidth]{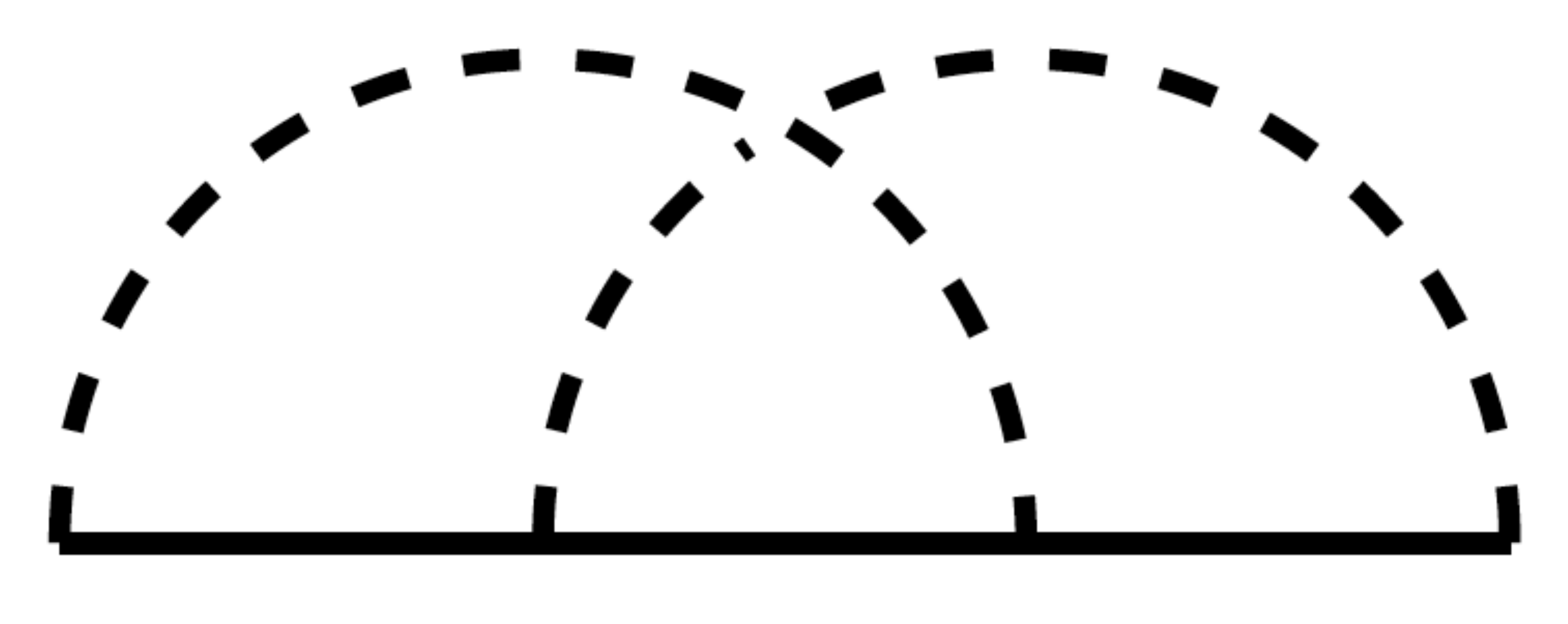}\quad
(d) \includegraphics[width=0.25\textwidth]{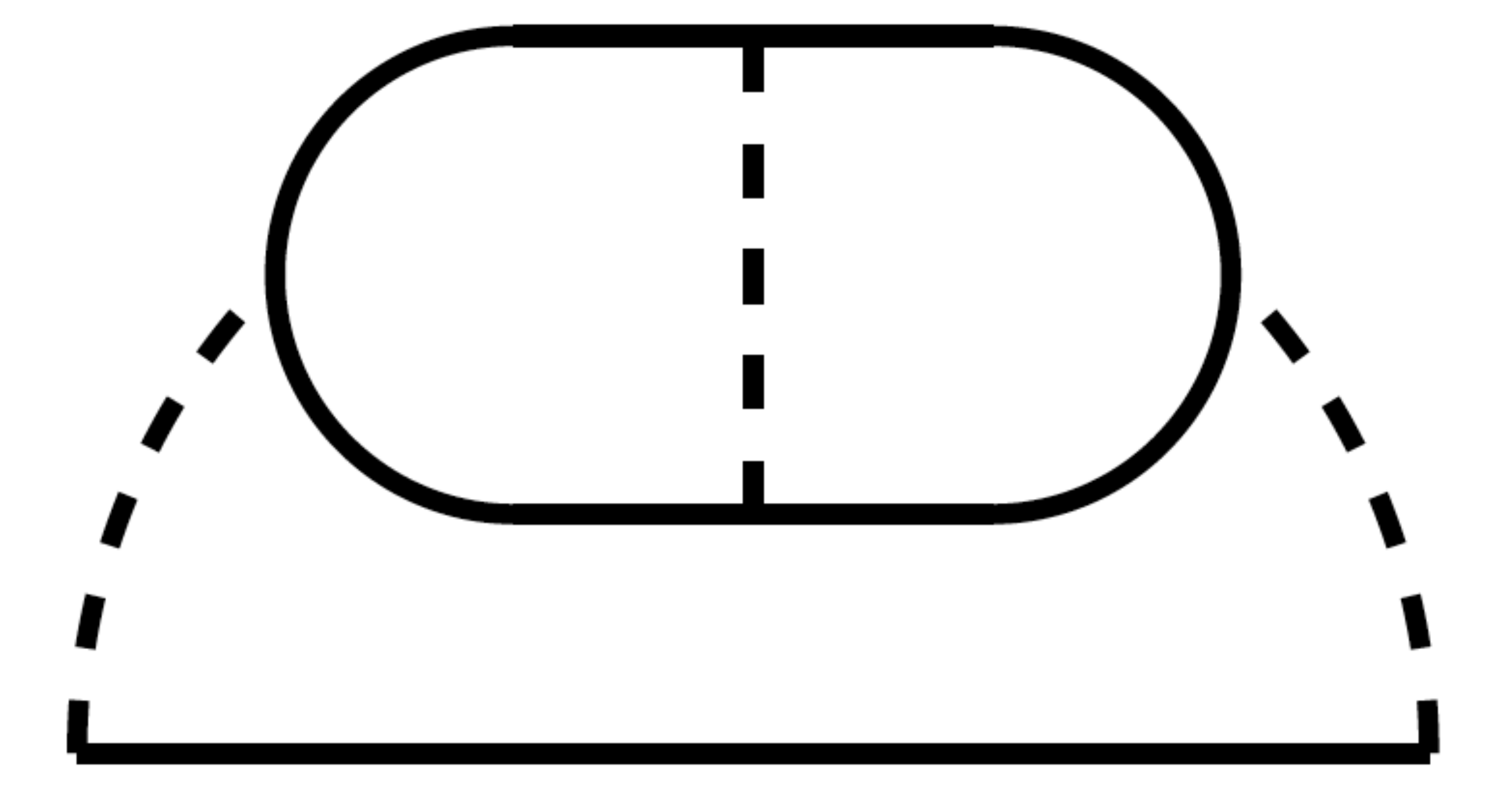}}
\centerline{(e) \includegraphics[width=0.25\textwidth]{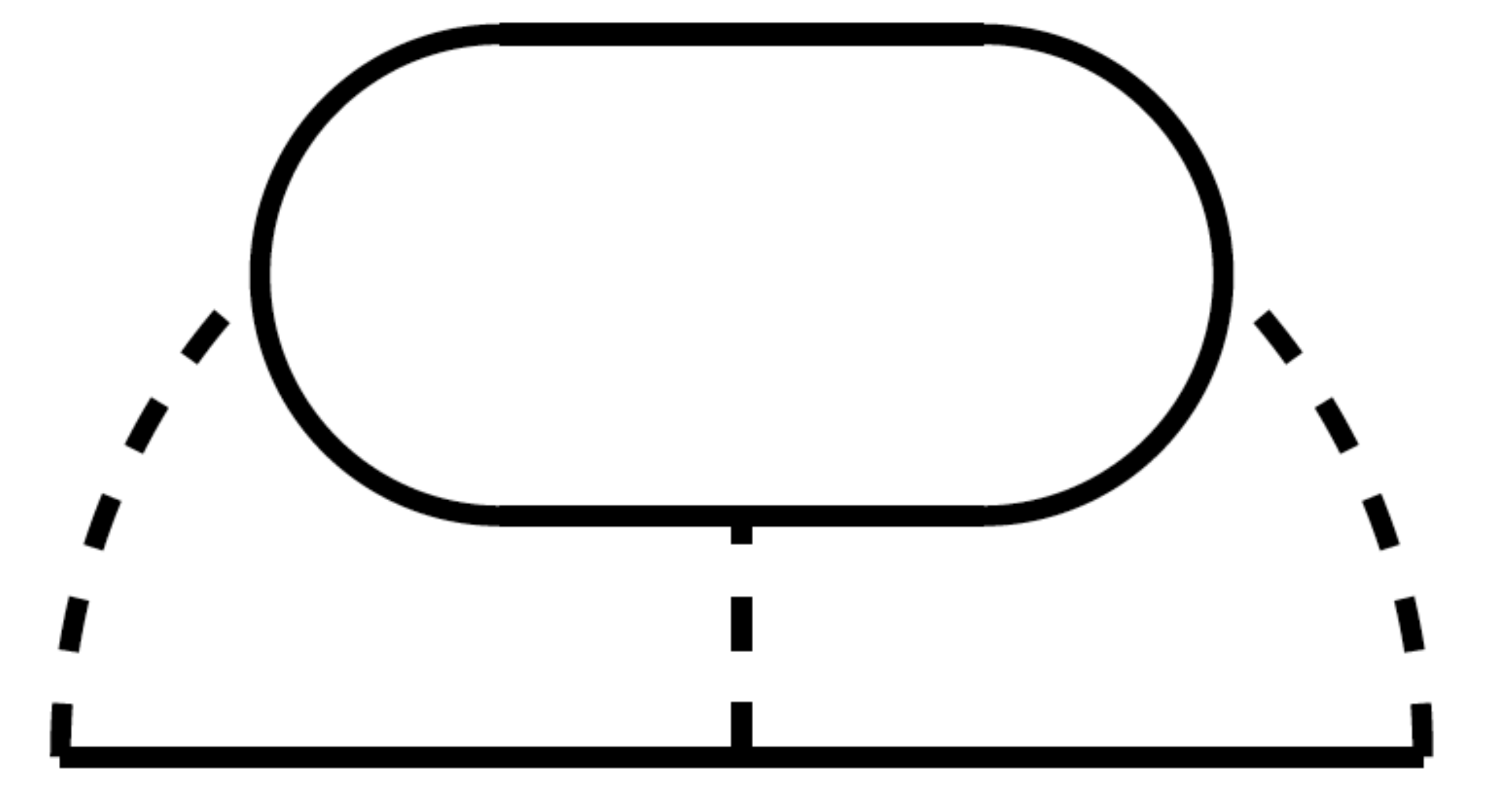}
\quad (f) \includegraphics[width=0.25\textwidth]{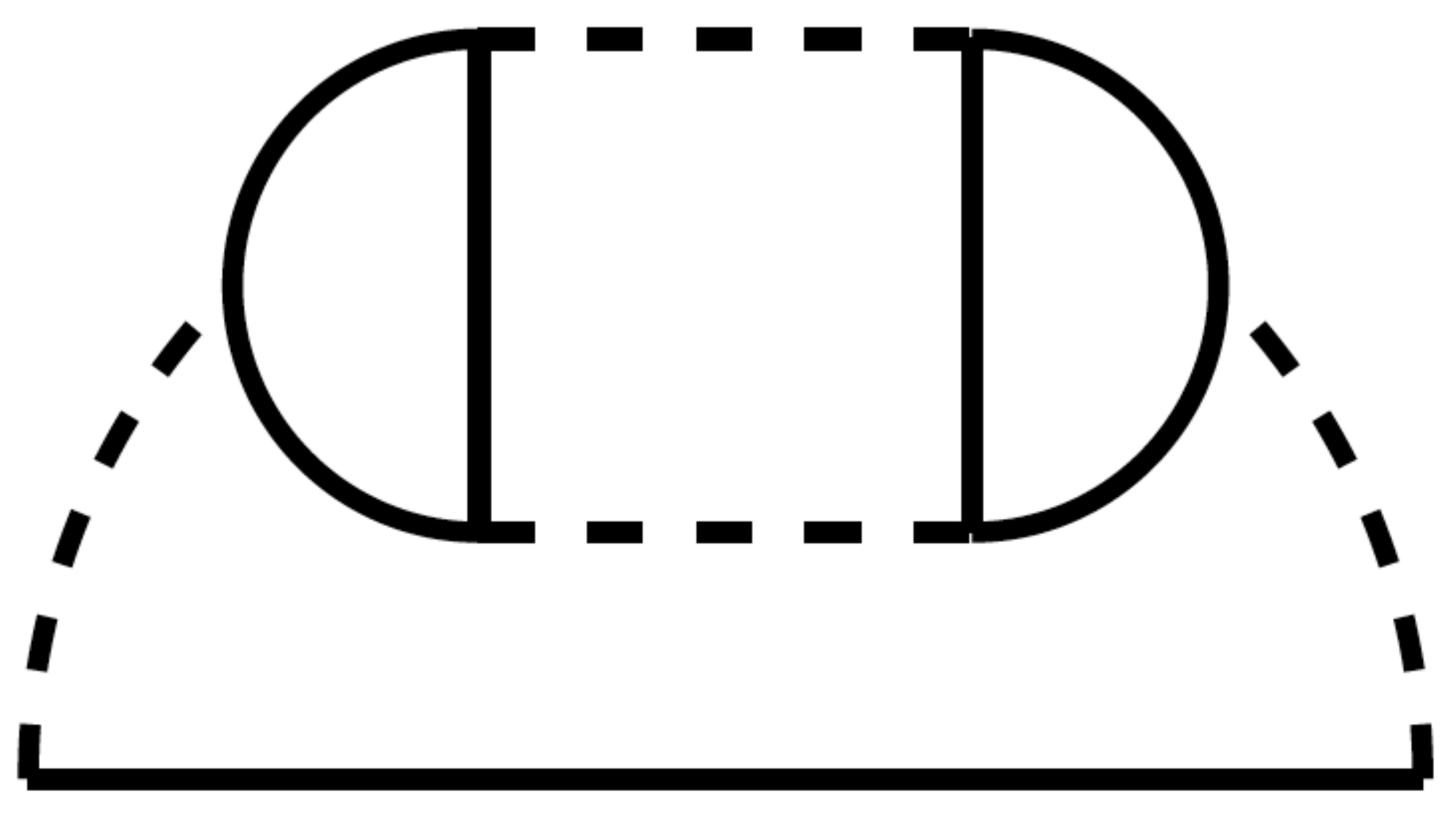}
}
\caption{Non-SCSA-type self-energy correction in the second order in $1/d_c$.}
\label{Fig3}
\end{figure}

\subsection{Non-SCSA-type corrections to $\eta$}

In addition to the SCSA type diagrams there exist four more diagrams for the self-energy in the second order in $1/d_c$ shown in Fig. \ref{Fig3}. Contrary to the SCSA type diagrams the diagrams in Fig. \ref{Fig3}c)-f) have only logarithmic divergence at the infrared. This allows us, on the one hand, to send $\tilde{q}_*$ to $\infty$  in the expressions for $N_q^{(0)}$, and, on the other, send the external momentum $k \to 0$ whenever it is possible. After such the procedure, we shall compute the integral by restoring the ultraviolet ($\tilde{q}_*$) and infrared cutoffs ($k$) in the most convenient way. However, contrary to the SCSA type diagrams, the evaluation of each of the diagrams in Fig. \ref{Fig3} is still involved. Details of the analytical evaluation of the integrals determining the non-SCSA corrections to $\eta$ can be found in Appendices \ref{App4} and \ref{App3}.

\subsubsection{Evaluation of $\Sigma^{(2,c)}_0(k)$}

The correction to the self-energy shown in Fig. \ref{Fig3}c) can be written as
\begin{gather}
\Sigma^{(2,c)}_0(k) = \frac{4}{T}
\int_{q,Q} \frac{[\bm{k \times q}]^2}{q^2}
 \frac{[\bm{k \times Q}]^2}{q^2} \frac{[\bm{(k-q) \times Q}]^2}{Q^2} \frac{[\bm{(k-Q) \times q}]^2}{q^2}
 \mathcal{G}^{0}_{\bm{k-q}}\mathcal{G}^{0}_{\bm{k-Q}}\mathcal{G}^{0}_{\bm{k-q-Q}}N^{(0)}_qN^{(0)}_Q  .
 \end{gather}
Taking the limit $\tilde{q}_*$ to $\infty$ and neglecting the external momentum $k$ in comparison with $q$ and $Q$, we find 
\begin{gather}
\Sigma^{(2,c)}_0(k) =  \left (\frac{32\pi}{3 d_c
}\right )^2\varkappa \int_{q,Q} \frac{[\bm{k \times q}]^2[\bm{k \times Q}]^2[\bm{Q \times q}]^4}{q^6Q^6|\bm{q-Q}|^4}
\end{gather}
Now since $\Sigma^{(2,c)}_0(k)$ depends only on the absolute value of $\bm{k}$, we can perform averaging of $\Sigma^{(2,c)}_0(k)$ over directions of $\bm{k}$. Then we find
 \begin{align}
\Sigma^{(2,c)}_0(k) &= \left (\frac{16\pi}{3 d_c
}\right )^2 \frac{\varkappa k^4}{2} \int_{q,Q} \frac{\bigl (q^2Q^2+(\bm{q\cdot Q})^2\bigr )[\bm{Q \times q}]^4}{q^6Q^6|\bm{q-Q}|^4}
\notag \\
& =\frac{32\varkappa k^4}{9 d_c^2} 
\int_0^\infty dq \int_0^\infty dQ 
\int_0^{2\pi}\frac{d\theta}{2\pi} \frac{q Q (1+2\cos^2\theta)\sin^4\theta}{(q^2+Q^2-2q Q\cos\theta)^2} .
\end{align}
Now using the following integral
\begin{equation}
\int_0^{2\pi}\frac{d\theta}{2\pi} \frac{(1+2\cos^2\theta)\sin^4\theta}{(q^2+Q^2-2q Q\cos\theta)^2}
=\frac{1}{16} \begin{cases}
(8q^2+5Q^2)/q^6 & Q\leqslant q, \\
(8Q^2+5q^2)/Q^6 & Q > q,
\end{cases}
\end{equation}
we find
 \begin{gather}
\Sigma^{(2,c)}_0(k) =\frac{4}{9 d_c^2} \varkappa k^4
 \int_0^\infty dq \int_0^q dQ 
\frac{q Q (8 q^2+5Q^2)}{q^6}=
\frac{7}{3 d_c^2}\varkappa k^4
\int_0^\infty \frac{dq}{q} \to \frac{7}{3 d_c^2}\varkappa k^4 \ln \frac{q_*}{k}
\label{eq:Sigma:2c}
\end{gather}

\subsubsection{Evaluation of $\Sigma^{(2,d)}_0(k)$}

The correction to the self-energy shown in Fig. \ref{Fig3}d) has the following form
\begin{gather}
\Sigma^{(2,d)}_0(k) = - \frac{4 d_c}{T^2}
\int_{Q,p,p^\prime} \frac{[\bm{k \times (p-p^\prime)}]^4}{|\bm{p-p^\prime}|^4}
 \frac{[\bm{p \times p^\prime}]^2}{|\bm{p-p^\prime}|^2} 
 \frac{[\bm{(p-Q) \times (p^\prime-Q)}]^2}{|\bm{p-p^\prime}|^2}
 \frac{[\bm{p \times Q}]^2}{Q^2} \frac{[\bm{p^\prime \times Q}]^2}{Q^2}\notag \\
 \times 
 \mathcal{G}^{(0)}_{\bm{k-p+p^\prime}}\mathcal{G}^{(0)}_p \mathcal{G}^{(0)}_{\bm{p-Q}}\mathcal{G}^{(0)}_{p^\prime} \mathcal{G}^{(0)}_{\bm{p^\prime-Q}}\Bigl[N^{(0)}_{|\bm{p-p^\prime}|}\Bigr ]^2N^{(0)}_Q 
 \end{gather}
Again taking the limit $\tilde{q}_*$ to $\infty$ and neglecting the external momentum $k$ in the argument of the Green's function, we find after averaging over directions of $\bm{k}$:
\begin{gather}
\Sigma^{(2,d)}_0(k) = - \left (\frac{16\pi}{3}\right )^3 \frac{3\varkappa k^4}{2 d_c^2} \int_{Q,p,p^\prime}
\frac{[\bm{p \times p^\prime}]^2[\bm{p \times Q}]^2[\bm{p^\prime \times Q}]^2[\bm{(p-Q) \times (p^\prime-Q)}]^2}{Q^2 p^4p^{\prime 4} |\bm{p-Q}|^4|\bm{p^\prime-Q}|^4|\bm{p-p^\prime}|^4} .
\end{gather}
Let us introduce the angles $\varphi=\angle (\bm{Q},\bm{p})$ and
$\varphi^\prime=\angle (\bm{Q},\bm{p^\prime})$. Then, we obtain
\begin{gather}
\Sigma^{(2,d)}_0(k) = - \left (\frac{16}{3 d_c}\right )^2 \varkappa k^4 \int_0^\infty d Q d p d p^\prime \int_0^{2\pi} \frac{d\varphi d\varphi^\prime}{(2\pi)^2}
\frac{p p^\prime Q^3 \sin^2\varphi\sin^2\varphi^\prime \sin^2(\varphi-\varphi^\prime)
}{(p^2+Q^2-2 p Q \sin\varphi)^2}
\notag \\
\times
\frac{
[p^\prime Q \sin\varphi^\prime -p Q \sin\varphi-pp^\prime\sin(\varphi^\prime-\varphi)]^2}{(p^{\prime 2}+Q^2-2 p^\prime Q \sin\varphi^\prime)^2 (p^2+p^{\prime 2}-2 p p^\prime \cos(\varphi-\varphi^\prime))^2}
\end{gather}
Let us make a change of variables $p \to p Q$ and $p^\prime \to p^\prime Q$, then we find
\begin{gather}
\Sigma^{(2,d)}_0(k) = \frac{b^{(2,d)}}{d_c^2} \varkappa k^4 \int_0^\infty \frac{dQ}{Q} \to  \frac{b^{(2,d)}}{d_c^2} \varkappa k^4  \ln \frac{q_*}{k} ,
\end{gather}
where 
\begin{gather}
b^{(2,d)} = - \left (\frac{16}{3}\right )^2 \int_0^\infty d p d p^\prime \int_0^{2\pi} \frac{d\varphi d\varphi^\prime}{(2\pi)^2}
\frac{p p^\prime \sin^2\varphi\sin^2\varphi^\prime \sin^2(\varphi-\varphi^\prime)
}{(p^2+1-2 p \sin\varphi)^2}
\notag \\
\times
\frac{
[p^\prime \sin\varphi^\prime -p  \sin\varphi-pp^\prime\sin(\varphi^\prime-\varphi)]^2}{(p^{\prime 2}+1-2 p^\prime \sin\varphi^\prime)^2 (p^2+p^{\prime 2}-2 p p^\prime \cos(\varphi-\varphi^\prime))^2} .
\label{b2d-int}
\end{gather}
We note that the expression under the integral signs is symmetric under the interchange of $p$ and $p^\prime$. The explicit calculation of the integral in Eq. (\ref{b2d-int}) presented in \ref{App4-1} yields 
$$b^{(2,d)}=-2.$$
Thus, we find
\begin{gather}
\Sigma^{(2,d)}_0(k) = - \frac{2}{d_c^2} \varkappa k^4  \ln \frac{q_*}{k}  .
\label{eq:Sigma:2d}
\end{gather}

\subsubsection{Evaluation of $\Sigma^{(2,e)}_0(k)$}

The correction to the self-energy shown in Fig. \ref{Fig3}e) has the following form
\begin{gather}
\Sigma^{(2,e)}_0(k) = - \frac{8 d_c}{T^2}
\int_{q,p,Q}
 \frac{[\bm{k \times q}]^2}{q^2}
 \frac{[\bm{k \times Q}]^2}{Q^2}
 \frac{[\bm{(k-q) \times (Q-q)}]^2}{|\bm{q-Q}|^2}
 \frac{[\bm{p \times q}]^2}{q^2} \frac{[\bm{p \times Q}]^2}{Q^2} 
 \notag \\
 \times
 \frac{[\bm{(p-Q) \times (p-q)}]^2}{|\bm{q-Q}|^2}
\mathcal{G}^{(0)}_{\bm{p-q}}\mathcal{G}^{(0)}_p \mathcal{G}^{(0)}_{|\bm{p-Q}|} \mathcal{G}^{(0)}_{|\bm{k-q}|}\mathcal{G}^{(0)}_{|\bm{k-Q}|}N^{(0)}_q N^{(0)}_Q 
N^{(0)}_{|\bm{q-Q}|} .
 \end{gather}
 Again we take the limit $\tilde{q}_*\to \infty$. Next neglecting the external momentum $k$ in the argument of the Green's functions, we find after averaging over directions of $\bm{k}$:
\begin{gather}
\Sigma^{(2,e)}_0(k) = - \left (\frac{16\pi}{3}\right )^3 \frac{\varkappa k^4}{d_c^2} \int_{Q,p,q}
\frac{(q^2Q^2+2(\bm{q\cdot Q})^2)[\bm{q \times Q}]^2
[\bm{p \times Q}]^2[\bm{p\times q}]^2[\bm{(p-Q) \times (p-q)}]^2}{Q^6 q^6 |\bm{q-Q}|^2 p^4 |\bm{p-Q}|^4|\bm{p-q}|^4} .
\end{gather}
Next we introduce angles $\varphi=\angle (\bm{p},\bm{q})$ and
$\varphi^\prime=\angle (\bm{p},\bm{Q})$. Then, we find
\begin{gather}
\Sigma^{(2,e)}_0(k) = - \left (\frac{8}{3 d_c}\right )^2 \frac{\varkappa k^4}{3} \int_0^\infty d p d Q  d q\int_0^{2\pi} \frac{d\varphi d\varphi^\prime}{(2\pi)^2}
\frac{p q Q \sin^2\varphi \sin^2\varphi^\prime (1+2\cos^2(\varphi-\varphi^\prime))}{(p^2+q^2-2 p q \cos\varphi)^2}
\notag \\
\times
\frac{\sin^2(\varphi-\varphi^\prime)[p Q \sin\varphi^\prime - p q \sin\varphi-q Q\sin(\varphi^\prime-\varphi)]^2}{(p^{2}+Q^2-2 p Q \cos\varphi^\prime)^2 (q^2+Q^{2}-2 q Q \cos(\varphi-\varphi^\prime))} .
\end{gather}
Let us change variables $Q \to p Q$ and $q \to p q$ then we find
\begin{gather}
\Sigma^{(2,e)}_0(k) = \frac{b^{(2,e)}}{d_c^2} \varkappa k^4 \int_0^\infty \frac{dp}{p} \to  \frac{b^{(2,e)}}{d_c^2} \varkappa k^4  \ln \frac{q_*}{k} ,
\end{gather}
where 
\begin{gather}
b^{(2,e)} = - \frac{64}{27} \int_0^\infty  d Q  d q\int_0^{2\pi} \frac{d\varphi d\varphi^\prime}{(2\pi)^2}
\frac{q Q \sin^2\varphi \sin^2\varphi^\prime (1+2\cos^2(\varphi-\varphi^\prime))}{(q^2+1-2 q  \cos\varphi)^2}
\notag \\
\times
\frac{\sin^2(\varphi-\varphi^\prime)[Q \sin\varphi^\prime - q \sin\varphi-q Q\sin(\varphi^\prime-\varphi)]^2}{(1+Q^2-2 Q \cos\varphi^\prime)^2 (q^2+Q^{2}-2 q Q \cos(\varphi-\varphi^\prime))} .
\label{b2e-int}
\end{gather}
We note that the expression under the integral sign is symmetric under the interchange $q$ and $Q$.
The calculation presented in  \ref{App4-2} yields
\begin{gather}
b^{(2,e)} = -\frac{58}{27} .
\end{gather}
Hence, we obtain
\begin{gather}
\Sigma^{(2,e)}_0(k) = - \frac{58}{27 d_c^2} \varkappa k^4  \ln \frac{q_*}{k} .
\label{eq:Sigma:2e}
\end{gather}

\subsubsection{Evaluation of $\Sigma^{(2,f)}_0(k)$}

The correction to the self-energy shown in Fig. \ref{Fig3}f) can be written as
\begin{gather}
\Sigma^{(2,f)}_0(k) = \frac{8 d^2_c}{T^3}
\int_{p,p^\prime,q,Q}
 \frac{[\bm{k \times q}]^4}{q^4} 
   \frac{[\bm{p \times q}]^2}{q^2} \frac{[\bm{p \times Q}]^2}{Q^2} 
 \frac{[\bm{(p-Q) \times (q-Q)}]^2}{|\bm{q-Q}|^2}
 \mathcal{G}^{(0)}_{|\bm{k-q}|}
 \mathcal{G}^{(0)}_p \mathcal{G}^{(0)}_{|\bm{p-Q}|} \mathcal{G}^{(0)}_{|\bm{p-q}|} 
 \notag \\
\times  \frac{[\bm{p^\prime \times q}]^2}{q^2} \frac{[\bm{p^\prime \times Q}]^2}{Q^2} 
 \frac{[\bm{(p^\prime-Q) \times (q-Q)}]^2}{|\bm{q-Q}|^2}
 \mathcal{G}^{(0)}_{p^\prime} \mathcal{G}^{(0)}_{|\bm{p^\prime-Q}|} \mathcal{G}^{(0)}_{|\bm{p^\prime-q}|} 
\Bigl[N^{(0)}_q\Bigr ]^2 N^{(0)}_Q 
N^{(0)}_{|\bm{q-Q}|} .
 \end{gather}
  Again we take the limit $\tilde{q}_*\to \infty$. Next neglecting the external momentum $k$ in the argument of the Green's functions, we find after averaging over directions of $\bm{k}$:
\begin{gather}
\Sigma^{(2,f)}_0(k) =  \left (\frac{16\pi}{3}\right )^4 \frac{3\varkappa k^4}{d_c^2} \int_{Q,q,p,p^\prime}
\frac{[\bm{p \times Q}]^2[\bm{p\times q}]^2[\bm{(p-Q) \times (q-Q)}]^2}{q^4 Q^2 |\bm{q-Q}|^2 p^4 |\bm{p-Q}|^4|\bm{p-q}|^4}
\notag \\
\times \frac{[\bm{p^\prime \times Q}]^2[\bm{p^\prime\times q}]^2[\bm{(p^\prime-Q) \times (q-Q)}]^2}{p^{\prime 4} |\bm{p^\prime-Q}|^4|\bm{p^\prime-q}|^4} .
\end{gather}
Now we introduce three angles: $\varphi=\angle (\bm{k},\bm{Q})$,
$\varphi^\prime=\angle (\bm{k^\prime},\bm{Q})$, and $\theta=\angle (\bm{q},\bm{Q})$. Then, we find
\begin{gather}
\Sigma^{(2,f)}_0(k) = \left (\frac{8}{3}\right )^4 \frac{3\varkappa k^4}{d_c^2} \int_0^\infty d p d p^\prime d Q  d q\int_0^{2\pi} \frac{d\varphi d\varphi^\prime d\theta}{(2\pi)^3}
\frac{p p^\prime}{q^3 Q (q^2+Q^2-2 q Q \cos\theta)}
\notag \\
\times 
 \frac{q^2 Q^2\sin^2\varphi \sin^2(\varphi-\theta)[p Q \sin\varphi-qQ \sin\theta -p q\sin(\varphi-\theta)]^2
 }{(p^2+Q^2-2p Q\cos\varphi)^2(p^2+q^2-2p q\cos(\varphi-\theta))^2}
 \notag \\
 \times
 \frac{q^2 Q^2\sin^2\varphi^\prime  \sin^2(\varphi^\prime -\theta)[p^\prime Q \sin\varphi^\prime -qQ \sin\theta -p^\prime q\sin(\varphi^\prime -\theta)]^2
 }{(p^{\prime 2}+Q^2-2p^\prime Q\cos\varphi)^2(p^{\prime 2}+q^2-2p^\prime q\cos(\varphi^\prime -\theta))^2} .
\end{gather}
Next we make a change of variables: $p = Q e^x$, $p^\prime= Q e^y$ and $q =Q e^z$. Then we find
\begin{gather}
\Sigma^{(2,f)}_0(k) = \frac{b^{(2,f)}}{d_c^2} \varkappa k^4 \int_0^\infty \frac{dQ}{Q} \to  
\frac{b^{(2,f)}}{d_c^2} \varkappa k^4  \ln \frac{q_*}{k} ,
\end{gather}
where 
\begin{gather}
b^{(2,f)} =\frac{3}{2} \left (\frac{8}{3}\right )^4  \int_{-\infty}^\infty d z \int_0^{2\pi} \frac{d\theta}{2\pi}
\frac{e^{-z} \Phi^2(z,\theta)}{\cosh z - \cos\theta} .
\label{b2f-int}
\end{gather}
Here the function $\Phi(z,\theta)$ is defined as follows
\begin{gather}
\Phi(z,\theta) = \frac{1}{16} e^z \int_{-\infty}^\infty d x \int_0^{2\pi} \frac{d\varphi}{2\pi}
\frac{\sin^2\varphi}{(\cosh x - \cos\varphi)^2}
\frac{\sin^2(\varphi-\theta)}{(\cosh (x-z) - \cos(\varphi-\theta))^2}
\notag \\
\times
\Bigl [e^{-z} \sin\varphi- e^{-x} \sin\theta -\sin(\varphi-\theta)\Bigr ]^2 .
\label{Phi-z-def}
 \end{gather}
We note that $\Phi(z,\theta)=\Phi(-z,\theta)=\Phi(z,-\theta)$. 

The integral in Eq.~(\ref{b2f-int}) is evaluated in \ref{App4-3}, yielding
\begin{gather}
b^{(2,f)} =
\frac{1}{9}+\frac{68\zeta(3)}{27}
\end{gather}
Hence, we find 
\begin{gather}
\Sigma^{(2,f)}_0(k) = \frac{3+68\zeta(3)}{27 d_c^2} \varkappa k^4  \ln \frac{q_*}{k} .
\label{eq:Sigma:2f}
\end{gather}

In total, using Eqs. \eqref{eq:Sigma:2c}, \eqref{eq:Sigma:2d}, \eqref{eq:Sigma:2e}, and \eqref{eq:Sigma:2d}, we obtain the contribution of the non-SCSA-type diagrams of Fig. \ref{Fig3} to the self-energy 
\begin{equation}
\Sigma^{(2,c)}_0(k)+\Sigma^{(2,d)}_0(k)+\Sigma^{(2,e)}_0(k)+\Sigma^{(2,f)}_0(k) 
= \frac{68\zeta(3)-46}{27 d_c^2}\varkappa k^4  \ln \frac{{q}_*}{k} .
\label{eq:Sigma:nSCSA}
\end{equation}

\subsection{Final result for $\eta$}

Now we can add up the contributions of the SCSA and non-SCSA types to the self energy upto the second order in $1/d_c^2$. Using Eqs. \eqref{eq:Sigma:SCSA} and \eqref{eq:Sigma:nSCSA}, we obtain 
\begin{gather}
\varkappa k^4- \Sigma^{(1)}_0(k) - \Sigma^{(2)}_0(k)
= \varkappa k^4 \Bigl [ 1 +\frac{2}{d_c} \ln \frac{q_*}{k} +\frac{2}{d_c^2} \ln^2 \frac{q_*}{k}+ \frac{73-68\zeta(3)}{27 d_c^2}\ln \frac{q_*}{k}\Bigr ] .
\label{eq:Sigma:full}
\end{gather}
As one can see, the non-SCSA-type diagrams have no smallness in comparison with the diagrams which are taken into account within SCSA. The result \eqref{eq:Sigma:full} translates into the result \eqref{eq:eta:main} for the bending rigidity exponent. The obtained result for $\eta$ implies that the value of the coefficient $b$ in Eq. \eqref{eq:nu:inter} is equal to 
\begin{equation}
b=(73-68\zeta(3))/27\approx -0.32. 
\label{final-b}
\end{equation}
Using this value we obtain the result \eqref{eq:nu:main} for the absolute Poisson's ratio.


\section{Conclusions\label{s4}}

To summarize, in this paper we studied a suspended 2D crystalline membrane embedded into a space of large dimensionality $d=2+d_c \gg 1$. We computed the absolute Poisson's ratio $\nu$ and the bending rigidity exponent $\eta$ to the second order in $1/d_c$. 

Our result \eqref{eq:nu:main} demonstrates that, for $\sigma_L \ll \sigma \ll \sigma_*$, the absolute Poisson's ratio of a 2D crystalline membrane is a universal but non-trivial function of $d_c$. Interestingly, the simple relation between the absolute Poisson's ratio and the exponent $\bm{\alpha}$, see Eq. \eqref{eq:c:1}, proposed in Ref. \cite{PR-PRB} breaks down at the order $1/d_c^2$ only. 

Our result \eqref{eq:eta:main} for the bending rigidity exponent $\eta$ indicates that, in agreement with general expectations, at each order of expansion in $1/d_c$ the non-SCSA-type diagrams provide the contribution of the same order as diagrams which are included into SCSA scheme. Therefore, the coincidence of $\eta^{\rm SCSA}$ at $d_c=1$ with numerical result for the bending rigidity exponents is a surprising occasion. 

Finally, we note that our results have been restricted to clean 2D membranes. It would be interesting to extend our analytical results for the $1/d_c$-expansion to the the case of a 2D disordered membrane.

\section{Acknowledgements}

We are grateful to A. Mirlin for useful discussions.  The work was funded in part by Deutsche Forschungsgemeinschaft, by the Alexander von Humboldt Foundation, by Russian Ministry of Science and Higher Educations, the Basic Research Program of HSE, and by Russian Foundation for Basic Research, grant No. 20-52-12019.

\newpage
\appendix

\section{The polarization operator in the presence of uniaxial stress \label{App1}}

In this Appendix we present details of analytical calculation of the polarization operator in the presence of the uniaxial stress, see Eq. \eqref{eq:Pi:def:0}. First step is to rewrite integral in real space representation
\begin{gather}
    \mathcal{P}(\bm{Q})= \int \frac{d^2\bm{k}}{(2\pi)^2}\frac{[\bm{k}\times\hat{\bm{Q}}]^2}{\bm{k}^4+k_x^2}\frac{[(\bm{Q}-\bm{k})\times\hat{\bm{Q}}]^2}{(\bm{Q}-\bm{k})^4+(Q_x-k_x)^2}
	= \int d^2\bm{x} \,e^{-i\bm{Q}\bm{x}} \left[ f(\bm{x},\bm{\hat Q}) \right]^2,
\end{gather}	
where $\bm{\hat Q} = \bm{Q}/Q$ and
\begin{gather}	
    f(\bm{x},\bm{\hat Q}) = \int\frac{d^2\bm{k}}{(2\pi)^2} \, e^{i\bm{k}\bm{x}}
 \frac{[\bm{k}\times{\bm{\hat Q}}]^2}{k^4+k^2_x}  .
  \end{gather}
Let us introduce the following notations
\begin{gather}
    f^{(a,b)}(\bm{x}) = \int\frac{d^2\bm{k}}{(2\pi)^2} \,  e^{i\bm{k}\bm{x}} \frac{k_x^ak_y^b}{k^4+k^2_x} ,
\end{gather}
where $a$ and $b$ are non-negative integers. Then we can write
\begin{equation}
f(\bm{x},\bm{\hat Q}) = \hat Q_x^2 f^{(0,2)}(\bm{x}) - 2  \hat Q_x \hat Q_y f^{(1,1)}(\bm{x})+\hat Q_y^2 f^{(2,0)}(\bm{x}) .
\end{equation}
In order to find the three functions $f^{(0,2)}$, $f^{(1,1)}$, and $f^{(2,0)}$, we first compute analytically
the functions $f^{(1,0)}$ and $f^{(2,0)}+f^{(0,2)}$:
\begin{gather}
    f^{(1,0)}(\bm{x}) = \sum_{\sigma =\pm} \frac{\sigma i}{2} \int \frac{d^2\bm{k}}{(2\pi)^2} \, \frac{e^{i\bm{k}\bm{x}}}{k^y + i \sigma k_x} 
                      = \sum_{\sigma =\pm} \frac{\sigma i}{2} e^{\sigma x/2} \int_0^\infty \frac{kdk}{2\pi}\frac{J_0(k\sqrt{x^2+y^2})}{k^2 + {1}/{4}} \\
                      = \frac{i}{2\pi} \sinh\left (\frac{x}{2}\right)  K_0\left (\frac{\sqrt{x^2+y^2}}{2}\right ) . \\
    f^{(2,0)}(\bm{x}) + f^{(0,2)}(\bm{x}) = \sum_{\sigma =\pm} \frac{1}{2} \int \frac{d^2\bm{k}}{(2\pi)^2} \,   \frac{e^{i\bm{k}\bm{x}}}{k^2 + i \sigma k_x} 
                      = \sum_{\sigma =\pm} \frac{1}{2} e^{\sigma x/2}  \int_0^\infty \frac{kdk}{2\pi}\frac{J_0(k\sqrt{x^2+y^2})}{k^2 + {1}/{4}}	\\
                      = \frac{1}{2\pi} \cosh \left (\frac{x}{2}\right ) K_0\left (\frac{\sqrt{x^2+y^2}}{2}\right ).
\end{gather}
Here $J_0(z)$ and $K_0(z)$ stands for the Bessel and modified Bessel functions of the zeroth order.
Next we use the relation $f^{(a+n,b+m)}(\bm{x})  = (-i\partial_x)^n(-i\partial_y)^m f^{(a,b)}(\bm{x})$ in order to find the three required functions:
\begin{gather}
    f^{(2,0)}(\bm{x}) = \frac{1}{4\pi} \left[\cosh\left (\frac{x}{2}\right ) K_0\left (\frac{\sqrt{x^2+y^2}}{2}\right ) - \frac{x}{\sqrt{x^2+y^2}}\sinh\left (\frac{x}{2}\right ) K_1\left (\frac{\sqrt{x^2+y^2}}{2}\right )\right], \notag	\\
    f^{(1,1)}(\bm{x}) = -\frac{1}{4\pi} \frac{y}{\sqrt{x^2+y^2}}\sinh\left (\frac{x}{2}\right ) K_1\left (\frac{\sqrt{x^2+y^2}}{2}\right ),	\notag \\
    f^{(0,2)}(\bm{x}) = \frac{1}{4\pi} \left[\cosh\left(\frac{x}{2}\right )K_0\left(\frac{\sqrt{x^2+y^2}}{2}\right ) + \frac{x}{\sqrt{x^2+y^2}}\sinh\left(\frac{x}{2}\right)K_1\left(\frac{\sqrt{x^2+y^2}}{2}\right)\right] .
\end{gather}

Next step is to perform inverse Fourier transform. We introduce the following notations $F^{(a,b;c,d)}(\bm{x}) = f^{(a,b)}(\bm{x}) f^{(c,d)}(\bm{x})$ and express the polarization operator as
\begin{gather}
    \mathcal{P}(\bm{Q})= \hat{Q}_x^4F^{(0,2;0,2)}(\bm{Q}) + 4\hat{Q}_x^2\hat{Q}_y^2F^{(1,1;1,1)}(\bm{Q}) + \hat{Q}_y^4F^{(0,2;0,2)}(\bm{Q}) 
        + 2 \hat{Q}_x^2\hat{Q}_y^2F^{(0,2,2,0)}(\bm{Q}) \notag \\- 4\hat{Q}_x^3\hat{Q}_yF^{(0,2,1,1)}(\bm{Q}) - 4\hat{Q}_x\hat{Q}_y^3F^{(2,0,1,1)}(\bm{Q}) ,
 \label{app:eq:pi_q}
\end{gather}
where $F^{(a,b;c,d)}(\bm{Q}) = \int d^2\bm{x} \exp(-i \bm{Q x}) F^{(a,b;c,d)}(\bm{x})$. In order to evaluate the functions $F^{(a,b;c,d)}(\bm{Q})$, we shall use the following identities:
\begin{align}
    \int_0^\infty dx\, x K_0^2\left(\frac{x}{2}\right)J_0(Qx) &= \frac{2\arcsinh Q}{Q\sqrt{1+Q^2}} \equiv 2A(Q),\label{eq:I1}\tag{I1}	\\
    \int_0^\infty dx\, x K_0\left(\frac{x}{2}\right)K_1\left(\frac{x}{2}\right)J_1(Qx) &= \frac{2\arcsinh Q}{\sqrt{1+Q^2}} \equiv 2  Q A(Q), \label{eq:I2}\tag{I2}\\
    \int_0^\infty dx \, K_1^2\left(\frac{x}{2}\right)\left[J_1(Qx)-\frac{Qx}{2}\right] &= 2\left[Q-\sqrt{1+Q^2}\arcsinh Q \right] . 
    \label{eq:I3}\tag{I3}
\end{align}

Below we demonstrate as an example the details of calculation of integrals for the function $F^{(1,1;1,1)}(\bm{Q})$:
\begin{gather}
    F^{(1,1;1,1)}(\bm{Q}) = \int \frac{d^2\bm{x}}{(4\pi)^2} e^{-i\bm{Q}\bm{x}} \left[ \frac{y}{\sqrt{x^2+y^2}}\sinh\left (\frac{x}{2}\right )K_1\left (\frac{\sqrt{x^2+y^2}}{2}\right )\right]^2	
        = \sum_{\sigma =\pm} \int \frac{d^2\bm{x}}{(8\pi)^2}e^{-i\bm{Q}\bm{x}} \notag \\
        \times
        \frac{y^2}{{x^2+y^2}}
        \bigl (e^{\sigma x}-1\bigr )K_1^2\left (\frac{\sqrt{x^2+y^2}}{2}\right ) 
        = \re \int_0^\infty \frac{dr}{16\pi} \frac{\partial}{\partial Q_y}\left[\hat{P}_yJ_1(Pr)-\hat{Q}_yJ_1(Q r)\right] K_1^2\left (\frac{r}{2}\right ) \notag	\\
        = \frac{1}{8\pi}\re\frac{\partial}{\partial Q_y} \left[Q_y(1+Q^2)A(Q) - P_y(1+P^2)A(P)\right]  ,
\end{gather}
where we introduced the vector $\bm{P} = (Q_1+i,Q_2)$. The other required functions are computed in a similar way. The results are as follows
\begin{align}
    8\pi F^{(0,2;0,2)}(\bm{Q}) &= \bigl (\hat{Q}_y^2 + Q^2 + 1\bigr ) A(Q) + \hat{Q}_x^2
        - \re\left [\bigl(\hat{P}_y^2 + P^2 - 1 + 2i P_x \bigr )A(P) + \hat{P}_x^2 \right ], \notag \\
    8\pi F^{(2,0;2,0)}(\bm{Q}) &= \bigl (\hat{Q}_y^2 + Q^2 + 1\bigr] A(Q) + \hat{Q}_x^2
            - \re\left [ \bigl( \hat{P}_y^2 + P^2 - 1 - 2iP_x\bigr)A(P) + \hat{P}_x^2 \right ], \notag \\
    8\pi F^{(1,1;1,1)}(\bm{Q}) &= \bigl(\hat{Q}_x^2 + Q^2\bigr )A(Q) + \hat{Q}_y^2
        - \re\left [ \bigl (\hat{P}_x^2 + P^2\bigr )A(P) + \hat{P}_y^2 \right ], \notag \\
   -8\pi F^{(0,2;2,0)}(\bm{Q}) &= \bigl (\hat{Q}_y^2 + Q^2 - 1\bigr ) A(Q) + \hat{Q}_x^2
            - \re\left [\bigl (\hat{P}_y^2 + P^2 + 1\bigr )A(P) + \hat{P}_x^2 \right ], \notag \\
   -8\pi F^{(0,2;1,1)}(\bm{Q}) &= \hat{Q}_x\hat{Q}_y\bigl [1-A(Q)\bigr ]
            - \re\left [\bigl (-\hat{P}_x\hat{P}_y + iP_y\bigr )A(P) + \hat{P}_x\hat{P}_y \right ], \notag \\
   -8\pi F^{(2,0;1,1)}(\bm{Q}) &= \hat{Q}_x\hat{Q}_y\bigl (A(Q)-1\bigr )
            - \re\left [ \bigl (\hat{P}_x\hat{P}_y + iP_y\bigr )A(P) - \hat{P}_x\hat{P}_y \right ] . 
\end{align}
That together with \eqref{app:eq:pi_q} sums up to the answer for the polarization operator, Eq. \eqref{eq:Pi:final}.

\section{Some details of numerical computation of the coefficient $c$ \label{App2}}

In this Appendix we present some details of numerical computation of the coefficient $c$ given by Eq. \eqref{eq:express:c}. It is convenient to write it as follows
\begin{equation}
c = c_x - c_y, \qquad c_\beta = 16 \pi \int \frac{d^2\bm{Q}}{(2\pi)^2}
\left (\frac{L^{(1)}_\beta(\bm{Q})}{\mathcal{P}(\bm{Q})} - 
\frac{L^{(0)}_{\beta}(\bm{Q})}{\mathcal{P}_0({Q})} \right ),
\end{equation}
where 
\begin{equation}
L^{(j)}_\beta(\bm{Q})= \int \frac{d^2\bm{K}}{(2\pi)^2} 
 \frac{[\bm{K}\times\bm{Q}]^4}{Q^4}
 \frac{K_\beta^2}{[K^4+K_x^2]^2[|\bm{K}-\bm{Q}|^4+j (K_x-Q_x)^2]} .
 \end{equation}
 We note that 
 \begin{equation}
 L_x^{(1)}(\bm{Q})=
 \frac{1}{2} \left [ \mathcal{P}(\bm{Q})+ \frac{1}{2} Q_x \frac{\partial \mathcal{P}(\bm{Q})}{\partial Q_x} + \frac{1}{2} Q_y \frac{\partial \mathcal{P}(\bm{Q})}{\partial Q_y} \right ] .
 \end{equation}
Also, we mention that similar to $ \mathcal{P}(\bm{Q})$ the functions  $L_x^{(1)}(\bm{Q})$ and $L_y^{(1)}(\bm{Q})$ has very anisotropic behaviour close to $Q=0$. This complicates numerical evaluation of the integrals. In order to circumvent this problem we make the following 
splitting $c = c_{x1}-c_{y1} + c_{xy2}$ where 
\begin{gather}
 c_{\beta 1}
 = 16 \pi \int \frac{d^2\bm{Q}}{(2\pi)^2}
\frac{L^{(1)}_\beta(\bm{Q}) [\mathcal{P}_0({Q})-\mathcal{P}(\bm{Q})]}{\mathcal{P}(\bm{Q})\mathcal{P}_0({Q})} 
\end{gather}
and
\begin{equation}
c_{xy2} = 16 \pi \int \frac{d^2\bm{Q}}{(2\pi)^2}
\frac{L^{(1)}_x(\bm{Q})-L^{(0)}_{x}(\bm{Q})-L^{(1)}_y(\bm{Q})+L^{(0)}_{y}(\bm{Q}) }{\mathcal{P}_0({Q})} .
\end{equation}
Numerical evaluation yields
\begin{equation}
c_{x1} = 2.819\pm 0.001, \qquad c_{y1} = 7.19 \pm 0.02.
\end{equation}
In order to evaluate $c_{xy2}$ it is convenient to use the following transformations
\begin{gather}
\int \frac{d^2\bm{Q}}{(2\pi)^2}
\frac{L^{(1)}_\beta(\bm{Q})-L^{(0)}_{\beta}(\bm{Q})}{\mathcal{P}_0({Q})} =
- \frac{16\pi}{3} \int \frac{d^2\bm{Q}}{(2\pi)^2}\frac{d^2\bm{K}}{(2\pi)^2} 
 \frac{[\bm{K}\times\bm{Q}]^4(K_x-Q_x)^2K_\beta^2}{Q^2[K^4+K_x^2]^2[|\bm{K}-\bm{Q}|^4+(K_x-Q_x)^2]|\bm{K}-\bm{Q}|^4}
 \notag \\
 = - \frac{16\pi}{3}\int \frac{d^2\bm{P}}{(2\pi)^2} \frac{d^2\bm{K}}{(2\pi)^2} 
 \frac{[\bm{K}\times\bm{P}]^4}{P^4 |\bm{K}-\bm{P}|^2}
 \frac{P_x^2K_\beta^2}{[K^4+K_x^2]^2[P^4+P_x^2]} .
\end{gather} 
Now introducing $K=P t$ and $P=\sqrt{y}$, we find that 
\begin{gather}
c_{xy2} 
= -\frac{8}{3\pi^2} \int_0^{2\pi} d\theta \int_0^{2\pi} d\varphi \int_0^\infty dt \int_0^\infty dy \, 
\frac{y t^3 \sin^4\theta \cos^2\varphi \cos[2(\theta+\varphi)]}
{[y t^2+\cos^2(\theta+\varphi)]^2[y+\cos^2\varphi]
[t^2+1-2 t \cos\theta]} .
\end{gather}
Now using the following result
\begin{equation}
\int_0^\infty dy
\frac{y}
{[y t^2+\cos^2(\theta+\varphi)]^2[y+\cos^2\varphi]
} = \frac{\cos^2(\theta+\varphi) - t^2 \cos^2\varphi +t^2 \cos^2\varphi 
\ln \frac{t^2 \cos^2\varphi}{\cos^2(\theta+\varphi)}}{t^2 [ \cos^2(\theta+\varphi) - t^2 \cos^2\varphi]^2} ,
\end{equation}
we obtain 
\begin{gather}
c_{xy2} 
= -\frac{8}{3\pi^2} \int_0^{2\pi} d\theta \int_0^{2\pi} d\varphi \int_0^\infty dt  \, 
\frac{t \sin^4\theta \cos^2\varphi \cos[2(\theta+\varphi)]}
{[t^2+1-2 t \cos\theta]} \notag \\
\times 
\frac{\bigl [\cos^2(\theta+\varphi) -t^2 \cos^2\varphi+ t^2 \cos^2\varphi 
\ln \frac{t^2 \cos^2\varphi}{\cos^2(\theta+\varphi)}\bigr ]}{[ \cos^2(\theta+\varphi) - t^2 \cos^2\varphi]^2} .
\end{gather}
We note that the integrand is symmetric under simultaneous transformation
$\theta\to 2\pi -\theta$ and $\varphi \to 2\pi - \varphi$. Therefore, we can rewrite the above integral as follows
\begin{gather}
c_{xy2} 
= -\frac{16}{3\pi^2} \int_0^{\pi} d\theta \int_0^{2\pi} d\varphi \int_0^\infty dt  \,
\frac{t \sin^4\theta \cos^2\varphi \cos[2(\theta+\varphi)]}
{[t^2+1-2 t \cos\theta]}\notag \\
\times 
\frac{\bigl [\cos^2(\theta+\varphi) -t^2 \cos^2\varphi+ t^2 \cos^2\varphi 
\ln \frac{t^2 \cos^2\varphi}{\cos^2(\theta+\varphi)}\bigr ]}{[ \cos^2(\theta+\varphi) - t^2 \cos^2\varphi]^2} .
\end{gather}
Numerical evaluation of this integral results in $c_{xy2}= 4.933\pm 0.001$. In total, we find 
$c = 0.56 \pm 0.02$.

\section{Evaluation of non-SCSA contributions to the self-energy \label{App4}}

In this Appendix, we present the detailed analytical calculation of the integrals involved in the non-SCSA contributions $\Sigma_0^{(2,d)}$, $\Sigma_0^{(2,e)}$, and $\Sigma_0^{(2,f)}$ to the self-energy.

\subsection{Evaluation of $b^{(2,d)}$ \label{App4-1}}

In this subsection, we calculate analytically the integral determining the coefficient $b^{(2,d)}_0(k)$ in Eq. (\ref{b2d-int}). Let us introduce the variables $x$ and $y$ such that
$p = \exp(x)$ and  $p^\prime=\exp(y)$. Also we introduce variables  $u=\exp(i\varphi)$ and $v=\exp(i\varphi^\prime) $. Then, we find
\begin{gather}
b^{(2,d)} = -\frac{2}{9}
\int_{-\infty}^\infty dx \int_{-\infty}^x dy \oint_{|v|=1} \frac{dv}{2\pi i v} 
 \oint_{|u|=1} \frac{du}{2\pi i u} \frac{(u^2-1)^2}{(u - e^x)^2(u - e^{-x})^2}
   \frac{(v^2-1)^2}{(v-e^y)^2(v-e^{-y})^2}
      \notag \\
  \times \frac{(v^2-u^2)^2 [e^{-x}(v-1/v)-e^{-y}(u-1/u)-(v/u-u/v)]^2}{(u-e^{x-y}v)^2(u-e^{-x+y}v)^2} .
\end{gather}
The positions of the poles in the expression under the integral signs in $u$ and $v$ complex planes
depend on values of $x$ and $y$. Therefore it is convenient to consider the three domains of integration over $x$ and $y$:
\begin{equation}
{\rm (I):}\, 0<y<x, \qquad  {\rm (II):} \, x>0\, \& y<0, \qquad 
{\rm (III):}\,  y<x<0 .
\label{eq:domains}
\end{equation}
For each domain we can determine the poles of the expression inside the unit circle, $|u|<1$:
\begin{equation}
{\rm (I):}\, u=e^{-x}, u=ve^{y-x}, \qquad 
{\rm (II):} u=e^{-x}, u=ve^{y-x}, \qquad 
{\rm (III):} u=e^{x}, u=ve^{y-x} .
\label{eq:poles:u:d}
\end{equation}
Also for all three domains there is a pole at $u=0$. Performing integration over $u$, we find that except the pole at $v=0$ the obtained expression has the following poles inside the unit circle, $|v|<1$:
\begin{equation}
{\rm (I):}\, v=e^{-y}, v=e^{y-2x}, \qquad 
{\rm (II):} \,  v=e^{y}, v=e^{y-2x}, \qquad 
{\rm (III):} \, v=e^{y} .
\label{eq:poles:v:d}
\end{equation}
After integration over $v$, we find
\begin{gather}
b^{(2,d)} = -\frac{4}{9}
\int_{0}^\infty dx \int_{0}^x dy
\frac{2 e^{2 (x+y)}+5 e^{4 (x+y)}-6 e^{4 x+2 y}+8 e^{6 x+2 y}-16 e^{2 x+4 y}+2 e^{4 x}-4 e^{6
   x}+9 e^{4 y}}{e^{2 (3 x+y)} \left(e^{2 x}-1\right)^2}
   \notag \\
    -\frac{4}{9}
\int_{0}^\infty dx \int_{-\infty}^0 dy
 \frac{2 e^{4 (x+y)}-16 e^{2 (x+y)}-6 e^{4 x+2 y}-4 e^{6 x+2 y}+2 e^{2 x+4 y}+5 e^{4 x}+8 e^{6 x}+9
   e^{4 y}}{e^{6 x-2 y} \left(e^{2 x}-e^{2 y}\right)^2}
\notag \\
   -\frac{4}{9}
\int_{-\infty}^0 dx \int_{-\infty}^x dy
\frac{9 e^{4 (x+y)}-6 e^{2 (x+y)}-16 e^{4 x+2 y}+2 e^{2 x+4 y}+8 e^{2 x}+5 e^{4 x}-4 e^{2 y}+2 e^{4
   y}}{e^{2x-2 y}\left(e^{2 y}-1\right)^2}  
   \notag \\
   = -\frac{2}{9} \left [\left (3-\frac{\pi^2}{6}\right )+\frac{11}{4} +\left (\frac{13}{4}+\frac{\pi^2}{6} \right )\right ] = -2
\end{gather}
This value yields Eq.~(\ref{eq:Sigma:2d}) of the main text.

\subsection{Evaluation of $b^{(2,e)}$ \label{App4-2}}

Here, we calculate the integral in Eq.~(\ref{b2e-int}) for coefficient $b^{(2,e)}$. 
We introduce the following variables $x= \ln q$, $y= \ln Q$, $u=\exp(i\varphi)$, and $v= \exp(i\varphi^\prime)$. Then, we find
\begin{gather}
b^{(2,e)} = \frac{4}{27}
\int_{-\infty}^\infty dx \int_{-\infty}^x dy \oint_{|v|=1} \frac{dv}{2\pi i v^2} 
 \oint_{|u|=1} \frac{du}{2\pi i u^2} \frac{(u^2-1)^2}{(u - e^x)^2(u - e^{-x})^2}
   \frac{(v^2-1)^2}{(v-e^y)^2(v-e^{-y})^2}
   \notag \\
  \times \frac{(v^2-u^2)^2[e^{-x}(v-1/v)-e^{-y}(u-1/u)-(v/u-u/v)]^2}{(u-e^{x-y}v)^2(u-e^{-x+y}v)} \left [1+\frac{1}{2}\left (\frac{u}{v}+\frac{v}{u}\right )^2\right ] .
\end{gather}
Again the pole structure of the expression under the integral signs in $u$ and $v$ complex planes
depend on values of $x$ and $y$. Therefore it is convenient to consider the three domains of integration over $x$ and $y$ defined in Eq. \eqref{eq:domains}. The poles inside the unit circle, $|u|<1$ are the same as for the diagram on Fig. \ref{Fig3}d), see Eq. \eqref{eq:poles:u:d}. 
Performing integration over $u$, one finds that the obtained expression has poles in $v$ inside the unit circle, $|v|<1$. Their positions are exactly the same as for the coefficient $b^{(2,d)}$, see Eq. \eqref{eq:poles:v:d}.
After integration over $v$, we find
\begin{gather}
b^{(2,e)} = \frac{4}{27}
\int_{0}^\infty dx \int_{0}^x dy
\frac{e^{-4 (2 x+y)} }{e^{2 x}-1}\Bigl (2 e^{4 (x+y)}-3 e^{6 (x+y)}+7 e^{4 x+2 y}-19 e^{6 x+2 y}+14 e^{8 x+2 y}+5 e^{6 x+4 y}\notag \\
 -23 e^{8
   x+4 y}  +e^{2 (x+5 y)}+19 e^{4 x+6 y}-7 e^{2 x+8 y}+5 e^{4 x+8 y}-14 e^{4 x}+21 e^{6 x}-5 e^{8 x}-3 e^{10
   y}\Bigr)
   \notag \\
    +\frac{4}{27}
    \int_{0}^\infty dx \int_{-\infty}^0 dy \, 
 e^{2 y-8 x} \Bigl(7 e^{2 (x+y)}+4 e^{4 (x+y)}-4 e^{4 x+2 y}+14 e^{6 x+2 y}+2 e^{2 x+4 y}-10 e^{4 x}
 \notag \\
  -28 e^{6 x}+3 e^{4
   y}\Bigr)
\notag \\
   +\frac{4}{27}
\int_{-\infty}^0 dx \int_{-\infty}^x dy
\frac{e^{2 y}}{1-e^{2 y}} \Bigl (-73 e^{4 (x+y)}-36 e^{6 (x+y)}+62 e^{4 x+2 y}-21 e^{6 x+2 y}-4 e^{4 y-2 x}+10 e^{2 x+4 y}\notag \\
 +59 e^{6
   x+4 y}-2 e^{6 y-4 x}  +2 e^{6 y-2 x}+2 e^{2 x+6 y}+23 e^{4 x+6 y}-28 e^{2 x}-10 e^{4 x}+14 e^{2 y}+2 e^{6
   y}\Bigr)
   \notag \\
   = \frac{4}{27} \left [\frac{2}{9} \bigl (8 - 3 \pi^2\bigr)-\frac{889}{144} +\left (\frac{2 \pi ^2}{3}-\frac{485}{48} \right )\right ] = -\frac{58}{27} .
\end{gather}
This results in Eq.~(\ref{eq:Sigma:2e}) of the main text.

\subsection{Evaluation of $b^{(2,f)}$ \label{App4-3}}

The evaluation of the integral in Eq.~(\ref{b2f-int}) for the coefficient $b^{(2,f)}$ turns out
to be most involved among the integrals determining the non-SCBA contributions to the self-energy (Fig. \ref{Fig3}). 
By introducing new variables $w=\exp(i\theta)$ and $u =\exp(i\varphi)$, we can write
\begin{gather}
b^{(2,f)}
= -2 \left (\frac{16}{3}\right )^3 \int_{0}^\infty d z \oint_{|w=1|} \frac{dw}{2\pi i}
\frac{\cosh z}{(w-e^z)(w-e^{-z})} \Phi^2(z,w) , \quad
\Phi(z,w)= \int_{-\infty}^\infty d x \, F(x,z,w), \notag \\
F(x,z,w) =-\frac{1}{64} e^z \oint_{|u=1|} \frac{du}{2\pi i u} \frac{(u^2-1)^2}{(u-e^x)^2(u-e^{-x})^2}
\frac{(u^2-w^2)^2}{(u-w e^{x-z})^2(u- w e^{z-x})^2}
\notag \\
\times
\Bigl[e^{-z} \bigl (u-u^{-1}\bigr)- e^{-x} \bigl(w-w^{-1}\bigr) -
\bigl(u w^{-1}-w u^{-1}\bigr)\Bigr]^2.
\end{gather}
Let us first, integrate over $u$ in the function $F(x,z,w)$ under assumption that $z>0$. The result of integration depends on the intervals in which $x$ is situated. We split it into three domains:
\begin{equation}
{\rm (I):}\,  x<0, \qquad 
{\rm (II):}\, 0\leqslant x <z, \qquad
{\rm(III):}\, z\leqslant x.
\end{equation}
Then pole structure inside the unit circle, $|u|<1$, can be summarized as follows
\begin{equation}
{\rm (I):}\,  u=e^{x}, u=w e^{x-z}, \quad 
{\rm (II):}\, u=e^{-x}, u=w e^{x-z}, \quad
{\rm (III):}\, u=e^{-x}, u=w e^{z-x} .
\end{equation}
Also there is always the pole at $u=0$. Then, we can write that $\Phi(z,w) = \Phi_1(z,w) + \Phi_2(z,w)$, where 
\begin{gather}
\Phi_1(z,w) = \int_{0}^z dx \, F_{II}(x,z,w), \qquad 
\Phi_2(z,w) = \int_0^\infty dx \, \Bigl [ F_{III}(x+z,z,w)+F_I(-x,z,w) 
\Bigr ] .
\end{gather}
The function $\Phi_1(z,w)$ can be found explicitly
\begin{gather}
\Phi_1(z,w) = \frac{e^{-4 z}}{128 w^2
   \left(e^z-w\right) \left(w- e^{-z}\right)}
\Biggl [ 3 w \left(w^2+1\right)^2-\left(w^2+1\right) e^z \left(3 w^4+2 w^2 (4
   z+5)+3\right)
   \notag\\
   -\left(w^2+1\right) e^{3 z} \left(w^4 (4 z-3)+2 w^2 (16 z-5)+4 z-3\right)+4 w e^{2 z} \left(w^4+\left(3
   w^4+8 w^2+3\right) z+1\right)
   \notag \\
   +w e^{4 z} \left(w^4 (8 z-7)+6 w^2 (4 z-1)+8 z-7\right) \Biggr ] ,
   \label{eq:Phi1zw}
   \end{gather}
whereas integration over $x$ in $\Phi_2(z,w)$ is not easy to perform analiticaly. Therefore, we rewrite the expression for $b^{(2,f)}$ in the following way
\begin{gather}
b^{(2,f)} = -2 \left (\frac{16}{3}\right )^3  \int_{0}^\infty d z \oint_{|w=1|} \frac{dw}{2\pi i}
\frac{\cosh z}{(w-e^z)(w-e^{-z})} [\Phi_1^2(z,w)
+2\Phi_1(z,w)\Phi_2(z,w)+\Phi_2^2(z,w)] \notag \\
\equiv b^{(2,f)}_{11}+2 b^{(2,f)}_{12}+ b^{(2,f)}_{22} .
\end{gather}
In order to compute $b^{(2,f)}_{11}$ we integrate over $w$. There are poles at $w=0$ and $w=e^{-z}$ inside the unit circle, $|w|<1$, in the $w$ complex plane. Then, we obtain the following result:
\begin{gather}
b^{(2,f)}_{11} =  \frac{1}{27} \int_{0}^\infty d z \frac{e ^{-9z}\cosh z}{(e^{2z}-1)^3}
\Biggl [9 -24 e^{2 z} (z+1)
+e^{4 z} \left(16 z^2+128 z-23\right)
 -8 e^{6 z} \left(16 z^2+34 z-13\right)\notag \\+e^{8 z} \left(368 z^2+192 z-21\right)-8 e^{10 z} \left(52 z^2-29 z+14\right)+e^{12
   z} \left(352 z^2-256 z+67\right)\Biggr ] .
\end{gather}
Finally, integrating over $z$, we find
\begin{gather}
b^{(2,f)}_{11} = \frac{20}{27}\zeta (3) +
\frac{13253}{58320} .
\end{gather}

Next we consider the contribution $b^{(2,f)}_{12}$. In order to proceed, we first integrate over $w$. 
Using the explicit expression \eqref{eq:Phi1zw} for the function $\Phi_1(z,w)$, we analyse the pole structure in the complex plane of $w$. Inside the unit circle $|w|<1$, there are poles at $w=0$, $w=e^{-z}$, and $w=e^{-2x-z}$. After integration over $w$, we obtain the following result:
\begin{gather}
b^{(2,f)}_{12} = \frac{2}{27} \int_{0}^\infty d z \int_0^\infty d x  \frac{e^{-8x-9z} \cosh z}{(e^{2z+2z}-1)^2}
\Biggl [ (134-436 z) e^{6 x+8 z}-21 e^{2 (x+z)}-6 e^{4 x+2 z}-12 e^{8 x+2 z} \notag \\
+(98-80 z) e^{8 (x+z)}+(20-40 z) e^{8
   x+6 z}+(6-12 z) e^{6 x+2 z}-4 (z-2) e^{10 x+6 z}+8 (z+13) e^{4 (x+z)}\notag \\
   +(8 z-169) e^{6 (x+z)}+2 (8 z-7) e^{8 x+4 z}+(8
   z-7) e^{10 x+8 z}-5 (8 z-7) e^{6 x+10 z}+3 (8 z-5) e^{6 x+4 z}
   \notag \\ 
   -4 (25 z+33) e^{2 x+4 z}+4 (44 z-37) e^{4 x+8 z}+4 (52
   z-23) e^{8 x+10 z}+(56 z-4) e^{10 (x+z)}+3 e^{10 x+4 z}   \notag \\
 -3 (64 z-51) e^{2 x+6 z}
  +(336 z+50) e^{4 x+6 z}+9 e^{6 x}+48 e^{2 z}+16 e^{4
   z} (4 z-3)
   \Biggr ] .
\end{gather}
Next, integrating over $x$, we obtain
\begin{gather}
b^{(2,f)}_{12} = \frac{1}{81} \int_{0}^\infty d z \, e^{-9 z} \cosh z 
\Biggl [27+48 z^2 \bigl ( e^{4 z}- e^{6 z} \bigr )+240 e^{8 z} z^2+336 e^{10 z} z^2-72 e^{2 z} z+44 e^{4 z} z
\notag \\
+404 e^{6 z} z-96 e^{8 z} z -228
   e^{10 z} z+12 e^{12 z} z+24 e^{2 z}-60 e^{4 z}-96 e^{6 z}+111 e^{8 z}-6 e^{10 z} \notag \\ -6 e^{2 z} \left(e^{4 z} (14-4
   z)+e^{10 z}+e^{2 z} (4 z+1)
   +e^{6 z} (20 z+6)+e^{8 z} (28 z-19)-3\right) \ln \left(e^z-1\right) \notag \\
   -6 e^{2 z} \left(e^{4
   z} (14-4 z)+e^{10 z}+e^{2 z} (4 z+1)+e^{6 z} (20 z+6)+e^{8 z} (28 z-19)-3\right) \ln \left(e^z+1\right) \Biggr ] .
\end{gather}
Finally, integrating over $z$ we find
\begin{gather}
b^{(2,f)}_{12} =\frac{4}{9} \zeta (3)+\frac{5789}{19440}- \frac{23}{972}  \pi ^2. 
\end{gather}

Now we turn our attention to the coefficient  $b^{(2,f)}_{22}$. Again, at first, we integrate over $w$.
Inside the unit circle, $|w|<1$, there are four poles: at $w=0$, $w=e^{-z}$, $w=e^{-2x-z}$, and $w=e^{-2y-z}$
(we took into account that $x,y,z>0$). Integrating over $w$, we  obtain the following result:
\begin{gather}
b^{(2,f)}_{22} = -2 \left (\frac{16}{3}\right )^3 \int_{0}^\infty d z \int_0^\infty d x \int_0^\infty d y\,  X_{22}(x,y,z)
\label{eq:X22:def}
\end{gather}
The resulting expression for $X_{22}(x,y,z)$ is rather cumbersome and we present it separately in  \ref{App3}. Next, after integration over $x$ and $y$, we find 
\begin{gather}
b^{(2,f)}_{22} = -\frac{1}{243} \int_{0}^\infty d z  \, \frac{e^{-9 z} \cosh z}{73728}
\Biggl [
e^{2 z} \left(-36 e^{2 z} \left(e^{2 z}+1\right)^2 \left(e^{2 z}-1\right)^3 \left( -\text{Li}_2\left(1-e^{-2
   z}\right) \right . \right .
   \notag \\ \left. 
   +\frac{1}{2} \left(\pi ^2-\ln ^2\left(1-e^{-2 z}\right)\right)-\frac{\pi ^2}{6}\right)  +216 e^{2 z} z^2-216
   e^{4 z} z^2-432 e^{6 z} z^2+432 e^{8 z} z^2
   +216 e^{10 z} z^2
   \notag\\ 
   -216 e^{12 z} z^2+13 e^{2 z}+179 e^{4 z}+427 e^{6 z}+351
   e^{8 z}
   +90 e^{10 z}+36 e^{12 z}-120 e^{2 z} z+804 e^{4 z} 
   \notag\\ 
   \left .
   +672 e^{6 z} z-1140 e^{8 z} z-264 e^{10 z} z+156 e^{12 z}
   z-108 z
   +144 e^{2 z} \left(e^{2 z}+1\right)^2 z \left(e^{2 z}-1\right)^3 \ln \left(e^{2 z}-1\right)\right)
   \notag \\
   +36 e^{4 z}
   \left(-9 e^{2 z}-12 e^{4 z}-12 e^{6 z}-9 e^{8 z}+e^{10 z}+1\right) \text{Li}_2\left(e^{-2 z}\right)
 +117 e^{2 z}-74
   e^{4 z}-406 e^{6 z}
   \notag\\ +1225 e^{8 z}+195 e^{10 z}-246 e^{12 z}-36 e^{14 z}-108 e^{2 z} z-120 e^{4 z} z+804 e^{6 z} z+672
   e^{8 z} z
   \notag \\
   -1140 e^{10 z} z-264 e^{12 z} z+156 e^{14 z} z-6 \pi ^2 e^{4 z}+6 \pi ^2 e^{6 z}+12 \pi ^2 e^{8 z}-12 \pi ^2
   e^{10 z}-6 \pi ^2 e^{12 z}\notag \\
+6 \pi ^2 e^{14 z}
      -90 e^{4 z} \left(e^{2 z}+1\right)^2 \left(e^{2 z}-1\right)^3 \ln
   ^2\left(e^{2 z}-1\right) +12 e^{2 z} \left(2 e^{6 z} (52-12 z)+48 e^{4 z}
   \right . \notag \\ 
   \left .
   +9 e^{8 z}+e^{2 z} (12 z-19)   +e^{10 z} (12
   z-13)-9\right) \left(e^{2 z}-1\right) \ln \left(e^{2 z}-1\right)+81
\Biggr ] .
\end{gather}
Here $\text{Li}_2(z) = \sum_{k=1}^\infty z^k/k^2$ stands for the polylogarithm. 
Finally, integrating over $z$, we find
\begin{gather}
b^{(2,f)}_{22} = 
\frac{8}{9} \zeta (3)-\frac{41507}{58320}+\frac{23}{486} \pi ^2 .
\end{gather}
Combing all contributions,  $b^{(2,f)}_{11}$, $b^{(2,f)}_{12}$, $b^{(2,f)}_{22}$, together, we obtain
\begin{gather}
b^{(2,f)} = b^{(2,f)}_{11}+b^{(2,f)}_{12}+b^{(2,f)}_{22}=
\frac{3+68\zeta(3)}{27},
\end{gather}
leading to Eq.~(\ref{eq:Sigma:2f}) of the main text.

\section{Expression for the function $X_{22}(x,y,z)$  \label{App3}}

In this Appendix we present the expression for the function $X_{22}(x,y,z)$ which is determined the right hand side of Eq. \eqref{eq:X22:def}. This function can be written as
\begin{equation}
X_{22} = 
\frac{e^{-8 x-8 y-9 z} \cosh z }{2048 \left(1-e^{2 (x+z)}\right) \left(1-e^{2 (y+z)}\right) \left(1-e^{2 (x+y+z)}\right)^5}
\Bigl [X^{(1)}_{22}(x,y,z) +X^{(2)}_{22}(x,y,z) \Bigr ] ,
\end{equation}
where
\begin{gather}
X_{22}^{(1)}(x,y,z) = 
9 e^{6 (x+y)}+48 e^{6 (x+z)}-48 e^{8 (x+z)}
+48 e^{6 (y+z)}-48 e^{8 (y+z)}
-32 e^{4
   (x+y+z)}+53 e^{6 (x+y+z)}
\notag \\
   +290 e^{8 (x+y+z)}-1081 e^{10 (x+y+z)}+1510 e^{12 (x+y+z)}
   +313 e^{14 (x+y+z)}+3 e^{18
   (x+y+z)}
\notag \\
   +107 e^{4 (2 x+y+z)}-21 e^{2 (3 x+y+z)}+107 e^{4 (x+2 y+z)}-90 e^{6 (2 x+2 y+z)}-185 e^{2 (7 x+7 y+6 z)}
\notag \\
   -21 e^{2 (x+3 y+z)}+48 e^{6
   x+2 z}+460 e^{6 (2 x+y+2 z)}+377 e^{8 (2 x+y+2 z)}-567 e^{4 (3 x+y+2 z)}-567 e^{4 (x+3 y+2 z)}
\notag \\
-6 e^{6 x+4 y+2 z}+48
   e^{6 y+2 z}-6 e^{4 x+6 y+2 z}+9 e^{6 x+6 y+2 z}-3 e^{8 x+6 y+2 z}-3 e^{6 x+8 y+2 z}-45 e^{8 x+8 y+2 z}
\notag \\
-15 e^{4 (2
   x+y+3 z)}-77 e^{4 (3 x+y+3 z)}+16 e^{4 (4 x+y+3 z)}+357 e^{2 (5 x+y+3 z)}-15 e^{4 (x+2 y+3 z)}+38 e^{6 (2 x+2 y+3
   z)}
\notag \\
-83 e^{6 (3 x+2 y+3 z)}+16 e^{4 (x+4 y+3 z)}+16 e^{4 (3 x+4 y+3 z)}-4 e^{4 (4 x+4 y+3 z)}+357 e^{2 (x+5 y+3 z)}+16
   e^{6 x+4 z}
\notag \\
   -80 e^{8 x+4 z}+167 e^{2 (5 x+y+4 z)}-123 e^{2 (6 x+y+4 z)}+36 e^{4 x+2 y+4 z}-112 e^{6 x+2 y+4 z}-169 e^{8
   x+2 y+4 z}
   \notag  \\
   +978 e^{4 (3 x+3 y+4 z)}+1260 e^{4 (4 x+3 y+4 z)}+167 e^{2 (x+5 y+4 z)}+16 e^{6 y+4 z}-123 e^{2 (x+6 y+4
   z)}-112 e^{2 x+6 y+4 z}
\notag \\
+63 e^{4 x+6 y+4 z}-9 e^{6 x+6 y+4 z}+21 e^{8 x+6 y+4 z}-80 e^{8 y+4 z}-169 e^{2 x+8 y+4 z}+21
   e^{6 x+8 y+4 z}-44 e^{8 x+8 y+4 z}
\notag \\
 +15 e^{10 x+8 y+4 z}+15 e^{8 x+10 y+4 z}+90 e^{10 x+10 y+4 z}+69 e^{2 (4 x+y+5
   z)}+158 e^{2 (5 x+y+5 z)}-183 e^{2 (6 x+y+5 z)}
\notag \\
-44 e^{2 (7 x+y+5 z)}+69 e^{2 (x+4 y+5 z)}+172 e^{4 (4 x+4 y+5 z)}-183
   e^{2 (x+6 y+5 z)}-481 e^{2 (5 x+6 y+5 z)}+370 e^{2 (6 x+6 y+5 z)}
\notag \\
+60 e^{2 (7 x+6 y+5 z)}-44 e^{2 (x+7 y+5 z)}-50 e^{2
   (5 x+7 y+5 z)}+60 e^{2 (6 x+7 y+5 z)}+35 e^{2 (7 x+7 y+5 z)}-15 e^{2 (8 x+7 y+5 z)}
\notag \\
-15 e^{2 (7 x+8 y+5 z)}-9 e^{2 (8
   x+8 y+5 z)}-48 e^{8 x+6 z}+16 e^{10 x+6 z}-116 e^{6 x+2 y+6 z}+128 e^{8 x+2 y+6 z}
\notag \\
-104 e^{4 x+4 y+6 z}-100 e^{6 x+4
   y+6 z}+387 e^{8 x+4 y+6 z}+143 e^{10 x+4 y+6 z}+1043 e^{2 (5 x+5 y+6 z)}
\notag \\
+753 e^{2 (6 x+5 y+6 z)}+1170 e^{2 (7 x+5 y+6
   z)}+16 e^{2 (8 x+5 y+6 z)}+1170 e^{2 (5 x+7 y+6 z)}-240 e^{2 (6 x+7 y+6 z)} 
   \notag \\
-39 e^{2 (8 x+7 y+6
   z)}-48 e^{8 y+6 z}+128 e^{2 x+8 y+6 z}+387 e^{4 x+8 y+6 z}+16 e^{2 (5 x+8 y+6 z)}
\notag \\
-328 e^{6 x+8 y+6 z}-39 e^{2 (7 x+8
   y+6 z)}+82 e^{8 x+8 y+6 z}+3 e^{2 (9 x+8 y+6 z)}-15 e^{10 x+8 y+6 z}+3 e^{2 (8 x+9 y+6 z)}
\notag \\
+16 e^{10 y+6 z}+143 e^{4
   x+10 y+6 z}-211 e^{6 x+10 y+6 z}-15 e^{8 x+10 y+6 z}+85 e^{10 x+10 y+6 z}-30 e^{12 x+10 y+6 z}
\notag \\
-30 e^{10 x+12 y+6
   z}-1162 e^{2 (5 x+5 y+7 z)}-1021 e^{2 (6 x+5 y+7 z)}+882 e^{2 (7 x+5 y+7 z)}-712 e^{2 (8 x+5 y+7 z)}
\notag \\
-2 e^{2 (9 x+5 y+7
   z)}-1021 e^{2 (5 x+6 y+7 z)}-318 e^{2 (6 x+6 y+7 z)}-2275 e^{2 (7 x+6 y+7 z)}+89 e^{2 (8 x+6 y+7 z)}
\notag \\
-2 e^{2 (9 x+6 y+7
   z)}-712 e^{2 (5 x+8 y+7 z)}+89 e^{2 (6 x+8 y+7 z)}+69 e^{2 (7 x+8 y+7 z)}+52 e^{2 (8 x+8 y+7 z)}+9 e^{2 (9 x+8 y+7
   z)}
\notag \\
-2 e^{2 (5 x+9 y+7 z)}-2 e^{2 (6 x+9 y+7 z)}-2 e^{2 (7 x+9 y+7 z)}+9 e^{2 (8 x+9 y+7 z)}-e^{2 (9 x+9 y+7 z)}+32
   e^{10 x+8 z}+16 e^{12 x+8 z}
\notag \\
-69 e^{6 x+2 y+8 z}-78 e^{8 x+2 y+8 z}+174 e^{6 x+4 y+8 z}+408 e^{8 x+4 y+8 z}-789 e^{10
   x+4 y+8 z}-58 e^{2 (5 x+5 y+8 z)}
\notag \\
+317 e^{2 (6 x+5 y+8 z)}+535 e^{2 (7 x+5 y+8 z)}-1062 e^{2 (8 x+5 y+8 z)}+87 e^{2 (9
   x+5 y+8 z)}-69 e^{2 x+6 y+8 z}
\notag \\
+174 e^{4 x+6 y+8 z}+317 e^{2 (5 x+6 y+8 z)}+560 e^{6 x+6 y+8 z}-731 e^{2 (7 x+6 y+8
   z)}+53 e^{8 x+6 y+8 z}-17 e^{2 (9 x+6 y+8 z)}
\notag \\
-272 e^{10 x+6 y+8 z}+149 e^{12 x+6 y+8 z}+535 e^{2 (5 x+7 y+8 z)}-731
   e^{2 (6 x+7 y+8 z)}+2390 e^{2 (7 x+7 y+8 z)}
\notag \\
-166 e^{2 (8 x+7 y+8 z)}-2 e^{2 (9 x+7 y+8 z)}+87 e^{2 (5 x+9 y+8 z)}-17
   e^{2 (6 x+9 y+8 z)}-2 e^{2 (7 x+9 y+8 z)}-15 e^{2 (8 x+9 y+8 z)}
\notag \\
-5 e^{2 (9 x+9 y+8 z)}+32 e^{10 y+8 z}-789 e^{4 x+10
   y+8 z}-272 e^{6 x+10 y+8 z}+644 e^{8 x+10 y+8 z}-257 e^{10 x+10 y+8 z}
\notag \\
-30 e^{12 x+10 y+8 z}+16 e^{12 y+8 z}+149 e^{6
   x+12 y+8 z}+185 e^{8 x+12 y+8 z}-30 e^{10 x+12 y+8 z}-80 e^{12 x+12 y+8 z}
\notag \\
+30 e^{14 x+12 y+8 z}+30 e^{12 x+14 y+8
   z}+45 e^{14 x+14 y+8 z}+105 e^{2 (7 x+5 y+9 z)}-126 e^{2 (8 x+5 y+9 z)}+21 e^{2 (9 x+5 y+9 z)}
     %
,
   \end{gather}
   \begin{gather}
X^{(2)}_{22}(x,y,z)=   -478 e^{2 (7 x+6 y+9
   z)}+571 e^{2 (8 x+6 y+9 z)}+105 e^{2 (5 x+7 y+9 z)}-478 e^{2 (6 x+7 y+9 z)}+616 e^{2 (7 x+7 y+9 z)}
\notag \\
-953 e^{2 (8 x+7
   y+9 z)}+19 e^{2 (9 x+7 y+9 z)}-126 e^{2 (5 x+8 y+9 z)}+571 e^{2 (6 x+8 y+9 z)}-953 e^{2 (7 x+8 y+9 z)}
\notag \\
+37 e^{2 (8 x+8
   y+9 z)}-19 e^{2 (9 x+8 y+9 z)}+15 e^{6 x+4 y+10 z}+185 e^{8 x+4 y+10 z}-199 e^{10 x+4 y+10 z}-28 e^{12 x+4 y+10 z}
\notag \\
+307
   e^{14 x+4 y+10 z}+15 e^{4 x+6 y+10 z}-102 e^{6 x+6 y+10 z}-846 e^{8 x+6 y+10 z}-321 e^{10 x+6 y+10 z}
\notag \\
   +1440 e^{12 x+6
   y+10 z}+273 e^{14 x+6 y+10 z}+185 e^{4 x+8 y+10 z}-846 e^{6 x+8 y+10 z}-1135 e^{8 x+8 y+10 z}
\notag \\
-106 e^{10 x+8 y+10
   z}-588 e^{12 x+8 y+10 z}-281 e^{14 x+8 y+10 z}-28 e^{4 x+12 y+10 z}+1440 e^{6 x+12 y+10 z}
\notag \\
-588 e^{8 x+12 y+10 z}+307
   e^{4 x+14 y+10 z}+273 e^{6 x+14 y+10 z}-281 e^{8 x+14 y+10 z}-291 e^{10 x+4 y+12 z}
\notag \\
+367 e^{14 x+4 y+12 z}-51 e^{8 x+6
   y+12 z}+47 e^{10 x+6 y+12 z}-598 e^{14 x+6 y+12 z}-273 e^{16 x+6 y+12 z}
\notag \\
-51 e^{6 x+8 y+12 z}+554 e^{8 x+8 y+12 z}+1539
   e^{10 x+8 y+12 z}-476 e^{12 x+8 y+12 z}-834 e^{14 x+8 y+12 z}
\notag \\
+163 e^{16 x+8 y+12 z}-291 e^{4 x+10 y+12 z}+47 e^{6 x+10
   y+12 z}+1539 e^{8 x+10 y+12 z}+367 e^{4 x+14 y+12 z}
\notag \\
-598 e^{6 x+14 y+12 z}-834 e^{8 x+14 y+12 z}-273 e^{6 x+16 y+12
   z}+163 e^{8 x+16 y+12 z}+75 e^{10 x+6 y+14 z}
\notag \\
+429 e^{12 x+6 y+14 z}-287 e^{14 x+6 y+14 z}-253 e^{16 x+6 y+14 z}+36
   e^{18 x+6 y+14 z}+18 e^{8 x+8 y+14 z}
\notag \\
-21 e^{10 x+8 y+14 z}-612 e^{12 x+8 y+14 z}+146 e^{14 x+8 y+14 z}+852 e^{16 x+8
   y+14 z}-23 e^{18 x+8 y+14 z}
\notag \\
+75 e^{6 x+10 y+14 z}-21 e^{8 x+10 y+14 z}+429 e^{6 x+12 y+14 z}-612 e^{8 x+12 y+14 z}-253
   e^{6 x+16 y+14 z}
\notag \\
+852 e^{8 x+16 y+14 z}+36 e^{6 x+18 y+14 z}-23 e^{8 x+18 y+14 z}-141 e^{12 x+8 y+16 z}-177 e^{14 x+8
   y+16 z}
\notag \\
-59 e^{18 x+8 y+16 z}-141 e^{8 x+12 y+16 z}-177 e^{8 x+14 y+16 z}-59 e^{8 x+18 y+16 z}+50 e^{14 x+14 y+20 z}
\notag \\
-70
   e^{16 x+14 y+20 z}+15 e^{18 x+14 y+20 z}-70 e^{14 x+16 y+20 z}
+14 e^{18 x+16 y+20 z}+15 e^{14 x+18 y+20 z}
\notag \\
+14 e^{16
   x+18 y+20 z}+23 e^{18 x+18 y+20 z}+36 e^{2 (x+2 (y+z))}+408 e^{4 (x+2 (y+z))}+460 e^{6 (x+2 (y+z))} 
\notag \\
+377 e^{8 (x+2
   (y+z))}-116 e^{2 (x+3 (y+z))}-77 e^{4 (x+3 (y+z))}-476 e^{4 (2 x+3 (y+z))}-83 e^{6 (2 x+3 (y+z))}
\notag \\
+16 e^{4 (4 x+3
   (y+z))}-78 e^{2 (x+4 (y+z))}+1260 e^{4 (3 x+4 (y+z))}+63 e^{6 x+4 (y+z)}+158 e^{2 (x+5 (y+z))}
\notag \\
-199 e^{2 (2 x+5
   (y+z))}-321 e^{2 (3 x+5 (y+z))}-106 e^{2 (4 x+5 (y+z))}-481 e^{2 (6 x+5 (y+z))}-50 e^{2 (7 x+5 (y+z))}
\notag \\
-100 e^{4 x+6
   (y+z)}+753 e^{2 (5 x+6 (y+z))}-240 e^{2 (7 x+6 (y+z))}-328 e^{8 x+6 (y+z)}-211 e^{10 x+6 (y+z)}
\notag \\
-287 e^{2 (3 x+7
   (y+z))}+146 e^{2 (4 x+7 (y+z))}+882 e^{2 (5 x+7 (y+z))}-2275 e^{2 (6 x+7 (y+z))}+69 e^{2 (8 x+7 (y+z))}
\notag \\
-2 e^{2 (9 x+7
   (y+z))}-1062 e^{2 (5 x+8 (y+z))}+53 e^{6 x+8 (y+z)}-166 e^{2 (7 x+8 (y+z))}-15 e^{2 (9 x+8 (y+z))}         
\notag \\
+644 e^{10 x+8
   (y+z)}
+185 e^{12 x+8 (y+z)}+21 e^{2 (5 x+9 (y+z))}+19 e^{2 (7 x+9 (y+z))}-19 e^{2 (8 x+9 (y+z))}\Biggr ]  .
\end{gather}

\vspace{1cm}


\begin{thebibliography}{49}


\bibitem{MerminWagner} N. D. Mermin, H. Wagner, \textit{Absence of ferromagnetism or antiferromagnetism in one-- or two--dimensional isotropic Heisenberg models}, Phys. Rev. Lett. 17 (1966) 1133. 



\bibitem{Hohenberg} P. C. Hohenberg, \textit{Existence of Long-Range Order in One and Two Dimensions}, Phys. Rev. 158 (1967) 383.

 \bibitem{AronovitzLubensky88} J. A. Aronovitz and T. C. Lubensky, \textit{Fluctuations of solid membranes}, Phys. Rev. Lett. 60 (1988) 2634.
 
 \bibitem{paczuski1988} M. Paczuski, M. Kardar, and D. R. Nelson, \textit{Landau Theory of the Crumpling Transition}, Phys. Rev. Lett. 60 (1988) 2638.

\bibitem{david1} F. David and  E. Guitter, \textit{Crumpling transition in elastic membranes: Renormalization group treatment}, Europhys.  Lett. 5 (1988) 709.

\bibitem{david2} E. Guitter, F. David, S. Leibler, and L. Peliti, \textit{Thermodynamical behavior of polymerized membranes}, J. Physique 50 (1989) 1787.

\bibitem{Doussal} P.\ Le Doussal and L.\ Radzihovsky, \textit{Self-consistent theory of polymerized membranes}, Phys. Rev. Lett 69 (1992) 1209.

\bibitem{Kownacki} J.-P. Kownacki, and D. Mouhanna, \textit{Crumpling transition and flat phase of polymerized phantom membranes}, Phys. Rev. E 79 (2009) 040101(R).

\bibitem{Gompper91} G. Gompper and D. M. Kroll, \textit{A polymerized membrane in confined geometry}, Europhys. Lett. {\bf 15}, 783 (1991).

\bibitem{Bowick96} M. J. Bowick, S. M. Catterall, M. Falcioni, G. Thorleifsson, and K. N. Anagnostopoulos, \textit{The flat phase of crystalline membranes}, J. Phys. I France {\bf 6}, 1321 (1996).


\bibitem{Los-PRB-2009} J. H. Los,  M. I. Katsnelson,  O. V. Yazyev,  K. V.  Zakharchenko, and A. Fasolino, \textit{Scaling properties of flexible membranes from atomistic simulations: Application to graphene}, Phys. Rev. B  {\bf 80}, 121405(R) (2009).

\bibitem{Troster} A. Tr\"oster, \textit{Fourier Monte Carlo simulation of crystalline membranes in the flat phase}, J. Physics: Conf. Series 454 (2013) 012032.

\bibitem{buck} E. Guitter, F. David, S. Leibler, and  L. Peliti, \textit{Crumpling and buckling transitions in polymerized membranes}, Phys. Rev. Lett. 61  (1988) 2949.

\bibitem{lower-cr-D2} J. Aronovitz, L. Golubovic, T. C. Lubensky, \textit{Fluctuations and lower critical
dimensions of crystalline membranes}, J. Physique, {\bf 50} (1989) 609.

\bibitem{nelson15} A. Kosmrlj and D. R. Nelson, \textit{Response of thermalized ribbons to pulling and bending}, Phys. Rev. B  93 (2016) 125431.


\bibitem{katsnelson16} J. H. Los, A. Fasolino, and M. I. Katsnelson,  \textit{Scaling behavior and strain dependence of in-plane elastic properties of graphene}, Phys. Rev. Lett. 116 (2016) 015901.

\bibitem{my-hooke} I. V.  Gornyi, V. Yu. Kachorovskii,  and A. D. Mirlin, \textit{Anomalous Hooke's law in disordered graphene}, 2D Materials 4 (2017)  011003.


\bibitem{nicholl15} R. J. T. Nicholl, H. J. Conley, N. V. Lavrik, I. Vlassiouk, Y. S. Puzyrev, V. P. Sreenivas, S. T. Pantelides, and K. I. Bolotin, \textit{The effect of intrinsic crumpling on the mechanics of free-standing graphene}, Nat. Comm. 6 (2015) 8789.

\bibitem{DoussalR} P.\ Le Doussal and L.\ Radzihovsky, \textit{Anomalous elasticity, fluctuations and disorder in elastic membranes}, Ann. Phys. 392 (2018) 340.

\bibitem{Falcioni1997} M. Falcioni, M. J. Bowick, E. Guitter, and G. Thorleifsson, \textit{The poisson ratio of crystalline surfaces}, Europhys.  Lett. (EPL), 38 (1997) 67.

\bibitem{Gazit2009} D. Gazit, \textit{Structure of physical crystalline membranes within the self-consistent screening approximation}, Phys. Rev. E 80 (2009) 041117.

\bibitem{Hasselmann} N. Hasselmann and F. L. Braghin, \textit{Nonlocal effective average action approach to crystalline phantom membranes}, Phys. Rev. E 83 (2011) 031137.

\bibitem{Mouhanna} O. Coquand and D. Mouhanna, \textit{The flat phase of quantum polymerized membranes}, Phys. Rev. E 94 (2016) 032125.

\bibitem{PR-PRB} I. S. Burmistrov, I. V. Gornyi, V. Yu. Kachorovskii, M. I. Katsnelson, J. H. Los, and A. D. Mirlin, \textit{Stress-controlled poisson ratio of a crystalline membrane: Application to graphene},
Phys. Rev. B 97 (2018) 125.

\bibitem{LandauLifshitz} L. D. Landau and E. M. Lifshitz, \textit{Course of Theoretical Physics, vol.7: Theory of Elasticity}, (Butterworth Heinemann, 1986)


\bibitem{dPR} I. S. Burmistrov, I. V. Gornyi, V. Yu. Kachorovskii, and A. D. Mirlin, \textit{Differential Poisson's ratio of a crystalline two-dimensional membrane}, Ann. Phys. 396 (2018) 119.


\bibitem{crump} I.V. Gornyi, V. Yu. Kachorovskii,  and A. D. Mirlin, \textit{Rippling and crumpling in disordered free-standing graphene}, Phys. Rev. B 92 (2015) 155428.

\bibitem{footnoteCrumpling} For details see discussion after Eq. (15) in Ref. \cite{crump}. The role of the term $\partial_\alpha \bm{u} \partial_\beta \bm{u}$ in $u_{\alpha\beta}$ for the crumpling transition at $d_c\gg 1$ will be discussed elsewhere (D.R. Saykin, I.V. Gornyi, V.Yu. Kachorovskii, I.S. Burmistrov, in preparation). 

\bibitem{Peliti1987} D.R. Nelson and L. Peliti, \textit{Fluctuations in membranes with crystalline and hexatic order}, J. Phys. (Paris) 48 (1987)  1085.

\end{thebibliography}
\end{document}